\documentclass[apj,tighten,iop]{emulateapj}
\usepackage{appendix,natbib}
\usepackage{amsmath}
\usepackage{courier}
\usepackage{booktabs}
\usepackage{upgreek}
\usepackage{lineno}
\usepackage[pdfpagemode=UseNone,pdfstartview=FitH,colorlinks=true,citecolor=blue,linkcolor=blue,urlcolor=blue]{hyperref}

\usepackage{xcolor}

\newcommand{\kmps}{\ensuremath{\mathrm{km~s}^{-1}}}

\accepted{for publication in ApJ}

\graphicspath{{./}{figures/}}

\shorttitle{Massive Compact Starburst Galaxies}
\shortauthors{Perrotta et al.}

\begin{document}

\title{Physical Properties of Massive Compact Starburst Galaxies with Extreme Outflows}

\author{Serena Perrotta$^{1\star}$, Erin R. George$^{1}$, Alison L. Coil$^{1}$, Christy A. Tremonti$^{2}$, David S.~N. Rupke$^{3}$, Julie D. Davis$^{2}$, Aleksandar M. Diamond-Stanic$^{4}$, James E. Geach$^{5}$, Ryan C. Hickox$^{6}$, John Moustakas$^{7}$, Grayson C. Petter$^{6}$, Gregory H. Rudnick$^{8}$, Paul H. Sell$^{9}$, Cameren Swiggum$^{2}$, Kelly E. Whalen$^{6}$}

\email{$^{\star}$s2perrotta@ucsd.edu}

\affil{$^{1}$Department of Astronomy, University of California, San Diego, CA 92092, USA  \\ 
$^{2}$Department of Astronomy, University of Wisconsin-Madison, Madison, WI 53706, USA \\
$^{3}$Department of Physics, Rhodes College, Memphis, TN, 38112, USA \\
$^{4}$Department of Physics and Astronomy, Bates College, Lewiston, ME, 04240, USA \\
$^{5}$Centre for Astrophysics Research, University of Hertfordshire, Hatfield, Hertfordshire AL10 9AB, UK\\
$^{6}$Department of Physics and Astronomy, Dartmouth College, Hanover, NH 03755, USA \\
$^{7}$Department of Physics and Astronomy, Siena College, Loudonville, NY 12211, USA \\
$^{8}$Department of Physics and Astronomy, University of Kansas, Lawrence, KS 66045, USA \\
$^{9}$Department of Astronomy, University of Florida, Gainesville, FL, 32611 USA\\
}

\begin{abstract}
    We present results on the nature of extreme ejective feedback episodes and the physical conditions of a population of massive ($\rm M_* \sim 10^{11} M_{\odot}$), compact starburst galaxies at z = $0.4-0.7$. We use data from Keck/NIRSPEC, SDSS, Gemini/GMOS, MMT, and Magellan/MagE to measure rest-frame optical and near-IR spectra of 14 starburst galaxies with extremely high star formation rate surface densities (mean $\rm \Sigma_{SFR} \sim 2000 \,M_{\odot} yr^{-1} kpc^{-2}$) and powerful galactic outflows (maximum speeds v$_{98} \sim$ $1000-3000$ \kmps). Our unique data set includes an ensemble of both emission (\ion{[O}{2]}$\lambda\lambda$3726,3729,  H$\beta$,  \ion{[O}{3]}$\lambda\lambda$4959,5007,  H$\alpha$, \ion{[N}{2]}$\lambda\lambda$6549, 6585, and \ion{[S}{2]}$\lambda\lambda$6716,6731) and absorption  (\ion{Mg}{2}$\lambda\lambda$2796,2803, and \ion{Fe}{2}$\lambda$2586) lines that allow us to investigate the kinematics of the cool gas phase (T$\sim$10$^4$ K) in the outflows.  Employing a suite of line ratio diagnostic diagrams, we find that the central starbursts are characterized by high electron densities (median n$_e \sim$ 530 cm$^{-3}$), and high metallicity (solar or super-solar). 
    We show that the outflows are most likely driven by stellar feedback emerging from the extreme central starburst, rather than by an AGN. We also present multiple intriguing observational signatures suggesting that these galaxies may have substantial Lyman continuum (LyC) photon leakage, including weak \ion{[S}{2]} nebular emission lines.
    Our results imply that these galaxies may be captured in a short-lived phase of extreme star formation and feedback where much of their gas is violently blown out by powerful outflows that open up channels for LyC photons to escape.

\end{abstract}


\section{Introduction}

Starburst galaxies represent a fundamental phase in galaxy evolution, as they are widely considered to be the transition stage between star-forming galaxies and massive, passively-evolving ellipticals \citep[e.g.,][]{cim08}. According to some theoretical scenarios \citep[e.g.,][]{dimat07,hop10}, this transition is initiated by highly dissipative major merger events \citep{san96}, producing strong bursts of star formation in very dense cores, and possibly triggering obscured black hole accretion. The starburst activity and subsequent black hole feedback can cause gas depletion and removal through powerful outflows \citep{san88, silk98}, leading to a passively evolving system \citep{kor92, spr05, hop08}. 

Observations have revealed that outflows are a common characteristic of star-forming galaxies over a broad range of masses and redshifts \citep[e.g.,][]{mar98, pet01, rub10, mar12, arr14, rub14, chi15, hec15, hec16, mcq19}. The incidence and properties of galactic outflows have been explored from z $\sim$\,0 \citep{che10} through z $\sim$ 0.5 $-$ 1.5 \citep{rub10,rub11,rub14} to the peak epoch of cosmic star formation at z $\sim$ 2 $-$ 3 with both absorption \citep{ste10} and emission lines \citep{str17,str18}. 
When multi-band observations of the same system are available, they show that outflows are multi-phase, having several co-spatial, possibly kinematically coherent components with a wide range in density and temperature \citep{hec17}. Galactic-scale outflows can be identified through different phase tracers: cold gas (e.g. $<$ 10$^4$ K) including molecules such as CO \citep{flu19, spi20} as well as neutral \ion{H}{1} and metals such as NaD \citep{hec00, che10, mar05, con17, bae18, rup18}. Cool gas (i.e., $\sim$ 10$^4$ K) including metal ion tracers such as \ion{Fe}{2}, \ion{Mg}{2}, \ion{[O}{3]} \citep{mar09, rub14}, \ion{C}{2}, \ion{[N}{2]} \citep{hec15, hec16}, \ion{Si}{2}, \ion{Si}{3}, \ion{Si}{4} and \ion{[O}{1]} \citep{chi15, hec15, hec16}, as well as H$\alpha$ \citep{sha09,cic16}. Warm gas (i.e., $\sim$ 10$^5$ K - 10$^6$ K) can also be traced by ionized metals such as \ion{N}{5}, \ion{O}{6} \citep{kac15, nie17}. Finally, hot gas (i.e., $>$ 10$^6$ K) probed with both hard and soft X-ray emission \citep{leh99, str04,str07, str09}.

While powerful outflows appear to be essential to rapidly shut off star formation, the physical drivers of this ejective feedback remain unclear. In particular, the relative role of feedback from stars versus supermassive black holes (SMBHs) in quenching star formation in massive galaxies is still widely debated \citep[e.g.,][]{hop12, gab14, wei18, kro20}. In this context, the observed correlations between outflow and host galaxy properties can provide some insight \citep{rub14, tan17}. For instance, considering galaxy samples with a wide dynamic range of intrinsic properties, the outflow velocity is found to scale with stellar mass (M$_*$), star formation rate (SFR), and SFR surface density ($\Sigma_{SFR}$). 
This suggests that the faster outflows tend to be hosted in massive galaxies with high and concentrated star formation \citep[e.g.,][]{tan17}, implying that the starburst phase could potentially drive impactful outflows. Studying galaxies with extreme physical conditions can provide constraints on astrophysical feedback processes.

Our team has been investigating a sample of galaxies at z = 0.4 - 0.8 initially selected from the Sloan Digital Sky Survey \citep[SDSS;][]{yor00} Data Release 8 \citep[DR8;][]{Aihara11} to have distinct signatures of young post-starburst galaxies. Their spectra are characterized by strong stellar Balmer absorption from B- and A-stars, and weak or absent nebular emission lines indicating minimal on-going star formation. They lie on the massive end of the stellar mass function (M$_*$ $\sim$ 10$^{11}$ M$_{\odot}$; \citealp{dia12}). Remarkably, the optical spectra of most of these objects exhibit evidence of ejective feedback traced by extremely blueshifted ($>$ 1000 \kmps) \ion{Mg}{2} $\lambda\lambda$2796,2803 interstellar absorption lines (\citealp{tre07}, Davis et al., in prep). The \ion{Mg}{2} kinematics imply galactic outflows much faster than the $\sim$500 \kmps ones typical of massive star-forming galaxies \citep{chi17}. This finding painted an interesting picture where these galaxies were thought to be post-starburst systems with powerful outflows that may have played a crucial role in quenching their star formation. Surprisingly,  many of these galaxies were detected in the Wide-field Infrared Survey Explorer \citep[WISE;][]{wri10}, and the modeling of their ultraviolet (UV) to near-IR spectral energy distribution (SED) suggested a high level of heavily obscured star formation ($>$ 50 M$_{\odot}$ yr$^{-1}$; \citealp{dia12}). Hubble Space Telescope (HST) imaging of 29 of these galaxies revealed they are extremely compact (R$_e$ $\sim$ few 100 pc). Moreover, these data showed complex morphologies with diffuse tidal features indicative of various major merger stages \citep{sel14, dia21}. Combining SFR estimates from WISE restframe mid-IR luminosities with physical size measures from HST imaging, we derived extraordinarily high $\Sigma_{SFR}$ $\sim$ 10$^{3}$ M$_{\odot}$ yr$^{-1}$ kpc$^{-2}$ \citep{dia12}, approaching the theoretical Eddington limit \citep{leh96, meu97, mur05, tho05}. 

These results led us to draw a new scenario where these starburst galaxies have a dense dusty star-forming core at the center of the galaxy, and a substantial part of their gas and dust is blown away by powerful outflows. In this context, the high $\Sigma_{SFR}$ may reasonably be the driver of the exceptionally fast gas outflows seen which, in turn, may be responsible for the onset of rapid star formation quenching. Millimeter data for two galaxies in our sample indicates that the molecular gas is being consumed by the starburst with exceptional efficiency \citep{gea13}, and expelled in an extended molecular outflow \citep{gea14}, leading to rapid gas depletion times. Interestingly, \citet{sel14} used a suite of multiwavelength observations to assess the AGN activity in a sub-sample of these starbursts, and found little evidence for current AGN activity in half of the sample ($<$ 10 per cent of the total bolometric luminosity), though past AGN episodes could not be ruled out. This finding is in line with stellar feedback being the main driver of the observed outflows. These compact starburst galaxies exhibit the fastest outflows ($>$ 1000 \kmps) and highest $\Sigma_{SFR}$ among star-forming galaxies at any redshift, therefore they are an exquisite laboratory to test the limits of stellar feedback. They could represent a brief but common phase of massive galaxy evolution.

Our team followed up one of these starburst galaxies (J2118, or Makani) with Keck Cosmic Web Imager \citep[KCWI;][]{mor18}. The data reveal a spectacular galactic outflow traced by \ion{[O}{2]} emission line, reaching far into the circumgalactic medium (CGM) of the galaxy \citep{rup19}. The \ion{[O}{2]} emission has a classic bipolar hourglass limb-brightened shape, and exhibits a complex structure: a larger-scale, slower outflow ($\sim$300 \kmps) and a smaller-scale, faster outflow ($\sim$1500 \kmps). The velocities and sizes of these two outflows map exactly to two previous starburst episodes that this galaxy experienced, detected through the rest-frame optical emission and inferred ages of stars in this galaxy. These outflows are therefore consistent with being formed during recent starburst episodes in this galaxy’s past. The KCWI data on Makani directly shows that galactic outflows feed the CGM, expelling gas far beyond the stars in galaxies. 

In this paper, we present new optical and near-IR observations for 14 of the most well-studied starburst galaxies in our sample. We use this in
combination with some ancillary data to characterize their extreme ejective feedback events and explore their potential role in quenching the star formation in the host systems. Our unique data set includes both emission and absorption lines that allow us to probe outflowing gas at different densities. We investigate both the nature of the outflows, as well as the physical conditions in the central dusty starburst.  We use an ensemble of line ratio diagrams as crucial diagnostics of gas ionization, electron density, and metallicity. 

The paper is organized as follows: Section \ref{section:Observations and Data Reduction} illustrates the sample selection, observations, and data reduction; Section \ref{sectio:Emission Line Fitting} describes our measurements of the emission line kinematics; Section \ref{section:Results} presents our main results in comparison to other relevant galaxy samples, and Section \ref{section:Discussion} discusses the more comprehensive implications of our analysis. Our conclusions are reviewed in Section \ref{section:Conclusions}.

Throughout the paper, we assume a standard $\Lambda$CDM cosmology, with H$_0$ = 70 \kmps Mpc$^{-1}$, $\Omega_m$ = 0.3, and $\Omega_{\Lambda}$ = 0.7. All spectra are converted to vacuum wavelengths and corrected for heliocentricity.

\section{Sample and Data Reduction}
\label{section:Observations and Data Reduction}
The parent sample for this analysis has been drawn from the SDSS as described 
by \citet{tre07, dia12, sel14, dia21} and Tremonti et al. (in prep). In brief, this sample contains 1198 galaxies at 0.35 $<$ z $<$ 1.0 with $i<20$ mag from the SDSS DR8, with post-starburst spectral features: B- or A-star dominated
stellar continua and moderately weak nebular emission. 
A sub-sample of 121/1198 galaxies with z $>$ 0.4 
(such that the \ion{Mg}{2}$\lambda\lambda$2796, 2803 doublet
is readily observable with optical spectrographs) has been the center of comprehensive follow-up observations, with the aim of constraining the physical mechanisms responsible for launching their energetic feedback. More details about the sample selection can be found in Davis et al. (in prep.) and Tremonti et al. (in prep.). We collected ground-based spectroscopy for 50 of these galaxies with the MMT/Blue Channel, Magellan/MagE, Keck/LRIS, Keck/HIRES and/or Keck/KCWI \citep{tre07, dia12, sel14, gea14, dia16, gea18, rup19}, X-ray imaging with $Chandra$ for 12/50 targets \citep{sel14}, radio continuum data with the NSF's Karl G. Jansky Very Large Array (JVLA/VLA) for 20/50 objects \citep{pet20}, and optical imaging with HST for 29/50 galaxies \citep[``HST sample''][]{dia12, sel14}. For the HST observations, we first focused on the 12 most AGN-like galaxies and then on the 17 galaxies with the youngest derived post-burst ages (t$_{burst}$ $<$ 300 Myr), yielding a sample of galaxies with bluer U-V colors and stronger emission lines than typically found in post-starburst samples. 
We also used multi-band HST imaging to investigate the physical conditions at the centers of the 12/29 galaxies with the largest SFR surface densities measured by \citet{dia12}, (30 M${\odot}$ yr$^{-1}$ kpc$^{-2}$ $<$ $\Sigma_{SFR}$ $<$ 2000 M${\odot}$ yr$^{-1}$ kpc$^{-2}$), and explored the young compact starburst component that makes them so extreme \citep{dia21}.

In this paper, we focus on 13 galaxies from the HST sample (6 from the 12/29 most AGN-like galaxies, and 7 from the 17/29 with the youngest post-burst ages) plus one additional target, J1622+3145, that shows clear signs of an outflow in its spectrum. The targets in our sample are listed in Table \ref{table1} along with some of their main properties (see Section \ref{subsection:properties}).

\begin{figure*}[htp!]
  \centering
   \includegraphics[width=0.85\textwidth]{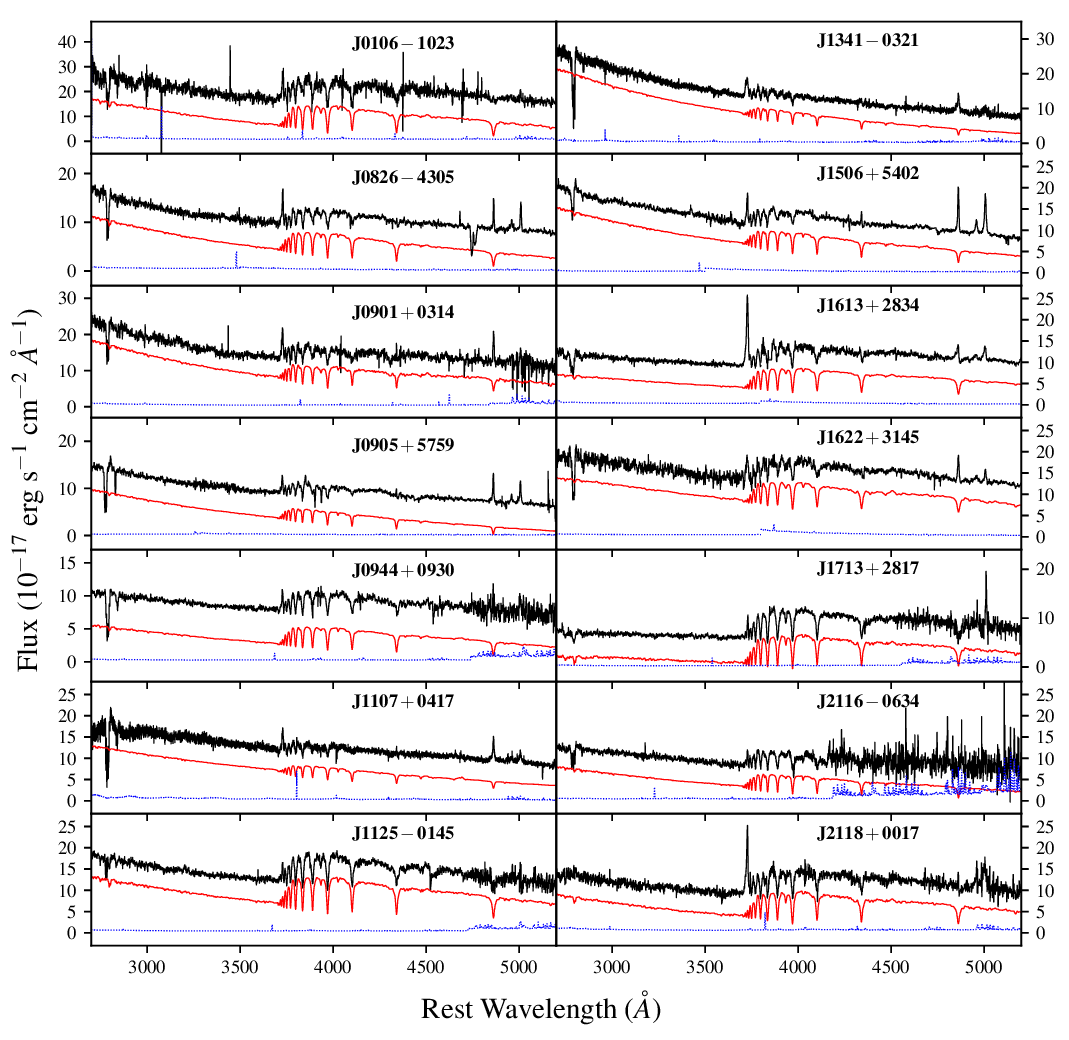}
  \caption{Rest-frame near-UV and optical spectra of the 14 galaxies in our sample. The black line shows the combined MMT, MagE and SDSS or GMOS spectra
(joined between 4500 and 4700 \AA, or 3500 and 3800 \AA). The red line represents the continuum model fit, offset in the vertical direction for clarity; errors from the best fit model are shown in blue. The
continuum model is subtracted from each spectrum before measuring the nebular emission lines of \ion{[O}{2]}$\lambda$3726, H$\beta$, and \ion{[O}{3]}$\lambda$5007. The spectra are dominated by the light of a young stellar population but have relatively weak nebular emission lines and strong \ion{Mg}{2} $\lambda \lambda$2796,2803 absorption originating from the interstellar medium.}
  \label{fig:Continuum}
\end{figure*}

\subsection{NIRSPEC}
\label{subsection:nirspec}

Near-IR spectra were obtained for the 13 targets selected from the HST sample, using the NIRSPEC cross-dispersed echelle spectrograph \citep{mcl98} on the Keck II telescope. Observing dates were September 15-17, 2013, and January 16-17, 2014. We used the NIRSPEC-1 filter covering 0.947-1.121 $\mu$m,
corresponding to the photometric Y band for the 11 sources at 0.45 $<$ z $<$ 0.68, and the NIRSPEC-2 filter covering 1.089-1.293 $\mu$m for the 2 sources at z $>$ 0.68. All targets were observed
with a 0.76 arcsec $\times$ 42 arcsecond slit with a spectral resolution of
R = $\lambda$/$\Delta \lambda$ $\approx$ 2000.  Individual exposures were 300 seconds, with total integration times of 40-60 minutes per object. We used the standard ABBA slit-nodding approach. We reduced the data using the REDSPEC IDL package \citep{kim15}. The exposures were dark subtracted and flat-fielded using an internal flat-field calibration lamp. We subtracted pairs of A–B exposures to perform sky subtraction. We performed relative flux calibrations and telluric absorption corrections using spectra of standard stars observed the same night.  
We determined the absolute flux calibration for the NIRSPEC spectra using the flux-calibrated MMT spectra available for each galaxy in our sample, as described in Section \ref{subsection:mmt}.

\begin{deluxetable*}{lrrrccccc}[htp!]
\tabletypesize{\small}
\tablecaption{Sample properties\label{table1}}
\tablehead{
\colhead{ID} & \colhead{$z$} & 
\colhead{RA} & \colhead{Dec.} & 
\colhead{log(M$_*$/M$_\odot$)} & \colhead{r$_e$} & 
\colhead{SFR} &
\colhead{$\Sigma_{SFR}$} &
\colhead{\ion{Mg}{2} Velocity} \\
\colhead{} & \colhead{} &
\colhead{(J2000)} & \colhead{(J2000)} &
\colhead{} & \colhead{(kpc)} & \colhead{(M$_\odot$ yr$^{-1}$)} &
\colhead{(M$_\odot$ yr$^{-1}$ kpc$^{-2}$)} & \colhead{(\kmps)} \\
\colhead{(1)} & \colhead{(2)} & \colhead{(3)} &
\colhead{(4)} & \colhead{(5)} & \colhead{(6)} &
\colhead{(7)} & \colhead{(8)} & \colhead{(9)}}
\startdata
J0106-1023 & 0.45 & 16.601056 & -10.391647& 10.72  & 0.590  & $166\substack{+33 \\ -31}$  &   76  & -1650 \\
J0826+4305 & 0.60 & 126.66006 & 43.091498 & 10.63  & 0.173  & $184\substack{+53 \\ -41}$  &  981 & -1425 \\
J0901+0314 & 0.46 & 135.38926 & 3.2367997 & 10.66  & 0.237  &  $99\substack{+39 \\ -26}$  &  281 & -1602 \\
J0905+5759 & 0.71 & 136.34832 & 57.986791 & 10.69  & 0.097  &  $90\substack{+23 \\ -20}$  & 1519  & -2910 \\
J0944+0930 & 0.51 & 146.07437 & 9.5053855 & 10.59  & 0.114  &  $88\substack{+26 \\ -21}$  & 1074  & -1679 \\
J1107+0417 & 0.47 & 166.76197 & 4.2840984 & 10.60  & 0.273  &  $73\substack{+13 \\ -14}$  &  155 & -2093 \\
J1125-0145 & 0.52 & 171.32874 & -1.7590066& 11.03  & 0.600  & $227\substack{+104 \\ -68}$  &   100  & -2309 \\
J1341-0321 & 0.66 & 205.40333 & -3.3570199& 10.53  & 0.117  & $151\substack{+34 \\ -23}$  & 1756  & -1936 \\
J1506+5402 & 0.61 & 226.65124 & 54.039095 & 10.60  & 0.168  & $116\substack{+32 \\ -25}$  & 652  & -2018 \\
J1613+2834 & 0.45 & 243.38552 & 28.570772 & 11.12  & 0.949  & $172\substack{+36 \\ -36}$ &   30  & -2699 \\
J1622+3145 & 0.44 & 245.69628 & 31.759132 & 10.62  & \nodata    & $151\substack{+52 \\ -33}$  & \nodata   & -1713 \\
J1713+2817 & 0.58 & 258.25161 & 28.285631 & 10.89  & 0.173  &  $229\substack{+99 \\ -72}$  &   1218 & -1298 \\
J2116-0624 & 0.73 & 319.10479 & -6.5791139& 10.41  & 0.284  & $110\substack{+55 \\ -27}$  &  216 & -2069 \\
J2118+0017 & 0.46 & 319.60026 & 0.2915070 & 10.95  & 2.240  & $230\substack{+93 \\ -76}$  &    5  & -1448 
\enddata
\tablecomments{-- Column 5: Stellar mass from Prospector. Column 6: Effective radii from HST. Column 7: SFRs from Prospector. Column 8: SFR surface densities estimated using columns (6) and (7). Column 9: \ion{Mg}{2}$\lambda$2796\,\AA\, maximum velocity, v$_{98}$.}
\end{deluxetable*}

\subsection{GMOS}
\label{subsection:gmos}
Five galaxies in our NIRSPEC sample (J0826+4305, J0905+5759, J1506+5402, J1613+2834, and J1713+2817) were also observed with Gemini Multi-Object Spectrographs \citep[GMOS;][]{all02, hoo04} on Gemini-North.
Here we use the GMOS data covering the H$\beta$ and \ion{[O}{3]} spectral region for these targets, and the H$\alpha$ one for J1613. We include in our final sample one additional target, J1622+3145, for which the GMOS spectrum covers the H$\alpha$ region and which shows unambiguous signs of an outflow.

The observations were carried out in service mode using Nod-and-Shuffle, spanning 16 nights from March 04, 2019 through April 23, 2019. A series of 360-seconds exposures were taken for each target, giving a total exposure time of $\sim$ 36 minutes.
The spectra were obtained with the NS0.75 arcsec long-slit, the Hamamatsu detector binned 2 $\times$ 2, and the R400$\_$G5305 grating, with a resulting spectral resolution of R $\approx$ 1920, and wavelength range from $\sim$ 0.36 to 1.03 $\mu$m. We adopted 0.745, 0.770 or 0.810 $\mu$m as central grating wavelength according to the redshift of the source, and spectrally dithered each pointing by $\pm$ 0.01 $\mu$m. This allows contiguous wavelength coverage in the presence of chip gaps and bad columns on the detector.

The data were reduced using the GMOS
sub-package in the Gemini PyRAF software package \citep[v1.14;][]{lab19}. Briefly, the data were bias subtracted and flat-fielded. The sky subtraction was performed subtracting the two shuffled sections of the detector. The GMOS data were then wavelength calibrated, extracted, and stacked. 
Relative flux calibration and telluric absorption correction were applied to the spectra based on standard stars observed at a similar airmass as the targets. We determined the absolute flux calibration of the GMOS data using the flux calibrated MMT spectra described in Section \ref{subsection:mmt}.

\subsection{ Other Optical Spectra}
\label{subsection:mmt}
  
We obtained the rest-frame UV–optical spectra of J1341 and J1107 with the Magellan Echellette (MagE) spectrograph \citep{mar08} on the Magellan Clay telescope with a 1 arcsec slit and 2 hours of integration time. The data were reduced and calibrated using the MASE pipeline \citep{boc09}. The spectra have a resolution R $\sim$ 4100 over a bandpass of 3300$-$9400 \AA\, in 15 orders ($\lambda_{rest} \sim 2300 - 6000$~\AA) and a signal-to-noise ratio (SNR) of $\sim$ 45 per resolution element near the galaxy’s \ion{Mg}{2}$\lambda\lambda$2796,2803 absorption lines. For all the other galaxies in our sample we collected high SNR optical spectra with the Blue Channel spectrograph on the 6.5m MMT between 2004 December and 2009 July \citep{tre07}. The data were obtained using a 1 arcsec long slit, which produced a FWHM resolution of 3.6~\AA\, (R $\sim$ 2000 near H$\beta$). The total exposure time for each target was $\sim$ 45-90 min. For our $z=0.4-0.8$ galaxies, this yielded rest-frame coverage from $\sim2700$ to 3900 \AA. The data were reduced, extracted, and spectrophotometrically calibrated using the ISPEC2D data reduction package \citep{mou06}.

There is extremely good agreement between the MMT, MagE, SDSS, and GMOS spectra where they overlap. We join the MMT, MagE, SDSS, and GMOS spectra when available, in order to extend our spectral coverage.
The combined spectra, including the stellar continuum fits, are shown in Fig.~\ref{fig:Continuum}. The systemic redshifts used throughout the paper are defined by the starlight.

The continuum model is built as described in \citet{gea18}. In brief, we fit the spectrum with a combination of simple stellar population (SSP) models and a \citet{cal00} reddening law. We employed the Flexible Stellar Population Synthesis code \citep{con09, con10} to generate SSPs with Padova 2008 isochrones, a \citet{sal55} initial mass function (IMF), and a theoretical stellar library “C3K” \citep{con18} with a resolution of R $\sim$ 10,000. We utilize solar metallicity SSP templates with 43 ages spanning 1 Myr$-$8.9 Gyr. We perform the fit with the Penalized Pixel-Fitting (pPXF) software \citep{cap04, cap17}. We mask forbidden emission lines and implement two separate templates for broad and narrow Balmer emission lines assuming Case B recombination line ratios. Both line and continuum are attenuated by the same amount of dust in the pPXF fit. By fitting Balmer emission and absorption lines simultaneously we can take into account the potential infill of the absorption line cores. One of the outputs of our pPXF fit is the stellar continuum model without any nebular component (shown in Fig.~\ref{fig:Continuum}).  We subtract from each spectrum our best fit pPXF model to properly remove the stellar component.  

Most sources, in addition to having strong Balmer absorption, show very blue continua indicating a recent starburst event ($\sim$ 1$-$10 Myr) that is not highly dust obscured.
These galaxies have morphologies of late-stage major mergers \citep{sel14}, which are consistent with 
having recent or on-going bursts of star formation. 
The MMT/MagE spectra allow high SNR measurements of the \ion{Mg}{2}$\lambda\lambda$2796,2803 interstellar medium (ISM) lines, used to search for signs of outflowing gas. \ion{Mg}{2} absorption lines are detected in all sources in our sample, with blueshifts with respect to the systemic redshift ranging from 1400 to 2900 \kmps. \citet{tre07} highlight the fact that these outflows are a factor of 2$-$5 times faster than the outflow velocities of typical IR-luminous star-forming galaxies (LIRGs and ULIRGs; e.g. \citep[LIRGs and ULIRGs; e.g.,][]{mar05, rup05}. We return to this point below in Section \ref{section:Results}.

\begin{deluxetable*}{lcccccc}[htp!]
 \centering
 \tabletypesize{\small}
 \tablecaption{Best Fit Parameters}
 \tablehead{
  \colhead{} & \colhead{H$\alpha$} & \colhead{H$\alpha$} &
  \colhead{H$\alpha$} & \colhead{\ion{[O}{2]}} & \colhead{\ion{[O}{2]}} & \colhead{\ion{[O}{2]}} \\
  \colhead{ID} & \colhead{Narrow FWHM} & 
  \colhead{Broad FWHM} & \colhead{$v_{off}$} & 
  \colhead{Narrow FWHM} & \colhead{Broad FWHM} & \colhead{$v_{off}$} \\
  \colhead{} & \colhead{(\kmps)} &
  \colhead{(\kmps)} & \colhead{(\kmps)}  & \colhead{(\kmps)}  & \colhead{(\kmps)}  & \colhead{(\kmps)} \\
  \colhead{(1)} & \colhead{(2)} & \colhead{(3)} &
  \colhead{(4)} & \colhead{(5)} & \colhead{(6)} & \colhead{(7)}}
  \startdata
  J0106-1023 & 525 $\pm$ 43 & \nodata & \nodata & 829 $\pm$ 39 & \nodata & \nodata \\
  
  J0826+4305 & 313 $\pm$ 33 &  918 $\pm$ 81 & -290 $\pm$ 56 & 414 $\pm$ 53 & 1761 $\pm$ 263 & -680 $\pm$ 171\\
  
  J0901+0314 & 410 $\pm$ 42 & \nodata & \nodata & 811 $\pm$ 30 & \nodata & \nodata \\
  
  J0905+5759 & 294$^{\dagger}$ $\pm$ 34 & 798$^{\dagger}$ $\pm$ 56 & -80$^{\dagger}$ $\pm$ 16 & 462 $\pm$ 77 & 1139 $\pm$ 175 & -380 $\pm$ 167 \\  
  
  J0944+0930 & 434 $\pm$ 61 & 1011 $\pm$ 345 & -67 $\pm$ 13  & 326 $\pm$ 128 & 925 $\pm$ 258 & -393 $\pm$ 249\\
  
  J1107+0417 & 481 $\pm$ 70 & 1985 $\pm$ 169 & -43 $\pm$ 9 & 451 $\pm$ 61 & 1534 $\pm$ 242 & 20 $\pm$ 8 \\
  
  J1125-0145 & 386 $\pm$ 43 & \nodata & \nodata & 417 $\pm$ 108 & 2396 $\pm$ 398 & -468 $\pm$ 174 \\

  J1341-0321 & 483 $\pm$ 35 & 1318 $\pm$ 132 & -205 $\pm$ 35 & 141 $\pm$ 29 & 1450 $\pm$ 25 & -262 $\pm$ 11 \\
  
  J1506+5402 & 358 $\pm$ 36 & 1218 $\pm$ 58 & -143 $\pm$ 25 & 523 $\pm$ 31 & 2058 $\pm$ 288 & -474 $\pm$ 158 \\
  
  J1613+2834 & 397 $\pm$ 56 & 1237 $\pm$ 65 & -257 $\pm$ 79 & 617 $\pm$ 25 & 1710 $\pm$ 68 & -308 $\pm$ 37 \\
 
  J1622+3145 & 482 $\pm$ 48 & 1071 $\pm$ 185 & -102 $\pm$ 37 & 415 $\pm$ 102 & \nodata & \nodata \\

  J1713+2817 & 521 $\pm$ 45 & \nodata & \nodata & 357 $\pm$ 78 & 1221 $\pm$ 551 & -577 $\pm$ 325 \\
  
  J2116-0624 & 112 $\pm$ 48 & 631 $\pm$ 85 & 15 $\pm$ 9 & 223 $\pm$ 89 & 1607 $\pm$ 420 & -245 $\pm$ 173 \\
  
  J2118+0017 & 281 $\pm$ 31 & 825 $\pm$ 45 & -231 $\pm$ 77 & 421 $\pm$ 42 & 1501 $\pm$ 84 & -341 $\pm$ 51 
  \enddata
 \tablecomments{-- Column 2-3: FWHMs of narrow and broad H$\alpha$ emission line components from NIRSPEC or GMOS spectra corrected for instrumental resolution. Column 4: velocity offset compared to systemic redshift of the broad H$\alpha$ component. Column 5-6: FWHMs of narrow and broad \ion{[O}{2]} emission line components from MMT, MagE or SDSS spectra corrected for instrumental resolution. Column 7: velocity offset compared to systemic redshift of the broad \ion{[O}{2]} component. $^{\dagger}$ We report values from the H$\beta$ emission line fit for J0905.\\}
 \label{table2}
\end{deluxetable*}
 
\subsection{Galaxy properties}
\label{subsection:properties}
Table \ref{table1} lists various relevant galaxy properties derived for sources in our sample.  
Stellar mass (M$_*$) and star formation rate (SFR) estimates are derived by fitting the broad-band UV -- mid-IR photometry and spectra with the Bayesian SED modelling code Prospector \citep{leja19, johnson21}, as described in Davis et al. (in prep). In brief, we include the 3500 - 4200 \AA\ spectral region in the fit since it contains many age-sensitive features (e.g., D4000, H$\delta$) and has a robust spectrophotometric calibration. SSP models are generated utilizing the Flexible Stellar Populations Synthesis code \citep[FSPS;][]{con09} assuming a Kroupa IMF \citep{kro01} and adopting the MIST isochrones \citep{cho16} and the C3K stellar theoretical libraries (Conroy et al., in prep.). The stellar models are very similar to the ones described in Section~\ref{subsection:mmt} over the wavelength range of interest for this work. 
The best fit parameters and their errors are calculated from the 16th, 50th, and 84th percentiles of the marginalized probability distribution function. See Davis et al. (in prep.) for examples of the SED fitting. The models fit the combined photometry and spectra well, however the lower SNR WISE W3 and W4 photometry and the limited infrared coverage of the SED provide poor constraints on the dust emission properties. This yields fairly tight constraints on the M$_*$ ($\pm$0.15 dex) and slightly larger errors on the SFR ($\pm$0.2 dex). M$_*$ represents the present day stellar mass of the galaxy and not the integral of the star formation history. In this work, we utilize SFRs computed from each galaxy's star formation history averaging over 100 Myr timescales. This is the characteristic timescale UV or IR star formation indicators are sensitive to \citep{ken12}. 


Measurements of the effective radii (r$_e$) for galaxies in our sample are discussed in \citet{dia12, sel14, dia21}. Briefly, for 3 galaxies (J0106, J1125, and J1713) we quantify the morphology using optical HST UVIS/F814W images. We employ GALFIT \citep{pen02, pen10} to model the two-dimensional surface brightness profile with a single Sersic component (defined by Sersic index $n$=4 and r$_e$), adopting an empirical model point-spread function (PSF) built using moderately bright stars in our science images. For the remaining 10 galaxies with multi-band imaging \citep{dia21}, we perform Serscic fits to the UVIS/F814W and UVIS/F475W images jointly using the GALFITM software \citep{haussler13, vika13}. To avoid uncertainties produced by tidal features, we fit the central region of the galaxy and  extrapolate the fit to larger radii to compute r$_e$.  The HST filters probe relatively blue ($\lambda_{rest}$(F475W) $\approx$ 3000\AA, $\lambda_{rest}$(F814W) $\approx$ 5200\AA) emission at z $\sim$ 0.6, tracing the young  unobscured stars rather than the stellar mass. Typical errors on the effective radius are of the order of 20\%. We do not have information on r$_e$ for one galaxy, J1622.

We also report maximum outflow velocities, derived from the \ion{Mg}{2}$\lambda\lambda$2796,2803 absorption lines  observed in MMT spectra, which show intricate velocity structures. We use VPFIT \citep[v10.4;][]{car14} to fit the doublet absorption profiles using a number of Voigt functions from one to six depending on the complexity of the lines. We parameterize the kinematics of \ion{Mg}{2} considering only one of the doublet components and measure the line velocity shift relative to the systemic redshift at which 98\% (v$_{98}$) of the equivalent width (EW) accumulates moving from red (positive velocities) to blue (negative velocities) across the line profile. The derived values in our sample range from -1400 to -2900 \kmps. To assess errors on v$_{98}$ due to uncertainties in the fits, we assume the best-fitting parameters are uncorrelated and vary them in a range of $\pm$1$\sigma$ and measure the resulting change in v$_{98}$. We use the largest variation of v$_{98}$ as upper limit error, with typical values of 200$-$400 \kmps for our sample.

\section{Emission Line Fitting}
\label{sectio:Emission Line Fitting}
 We quantify the kinematics of several diagnostic emission lines \ion{[O}{2]}$\lambda\lambda$3726,3729, H$\beta$, \ion{[O}{3]}$\lambda\lambda$4959,5007, H$\alpha$, \ion{[N}{2]}$\lambda\lambda$6549, 6585, and \ion{[S}{2]}$\lambda\lambda$6716,6731 for each galaxy in our sample as follows. After subtracting the best-fitting stellar population model of the galaxy (see Section~\ref{subsection:mmt}), the residual emission lines are fit using a custom Python algorithm. We model each emission line with one or two Gaussian functions, according to the complexity of the emission profiles and the SNR. A second Gaussian component is added only if the improvement in $\chi^2$ is statistically significant, accounting for the additional free parameters. Broadened or shifted emission line components trace gas with different kinematics from the rest of the ionized gas in the galaxy. Such components potentially trace outflowing gas.

\begin{figure*}[htp!]
  \centering
  \includegraphics[width=1\textwidth]{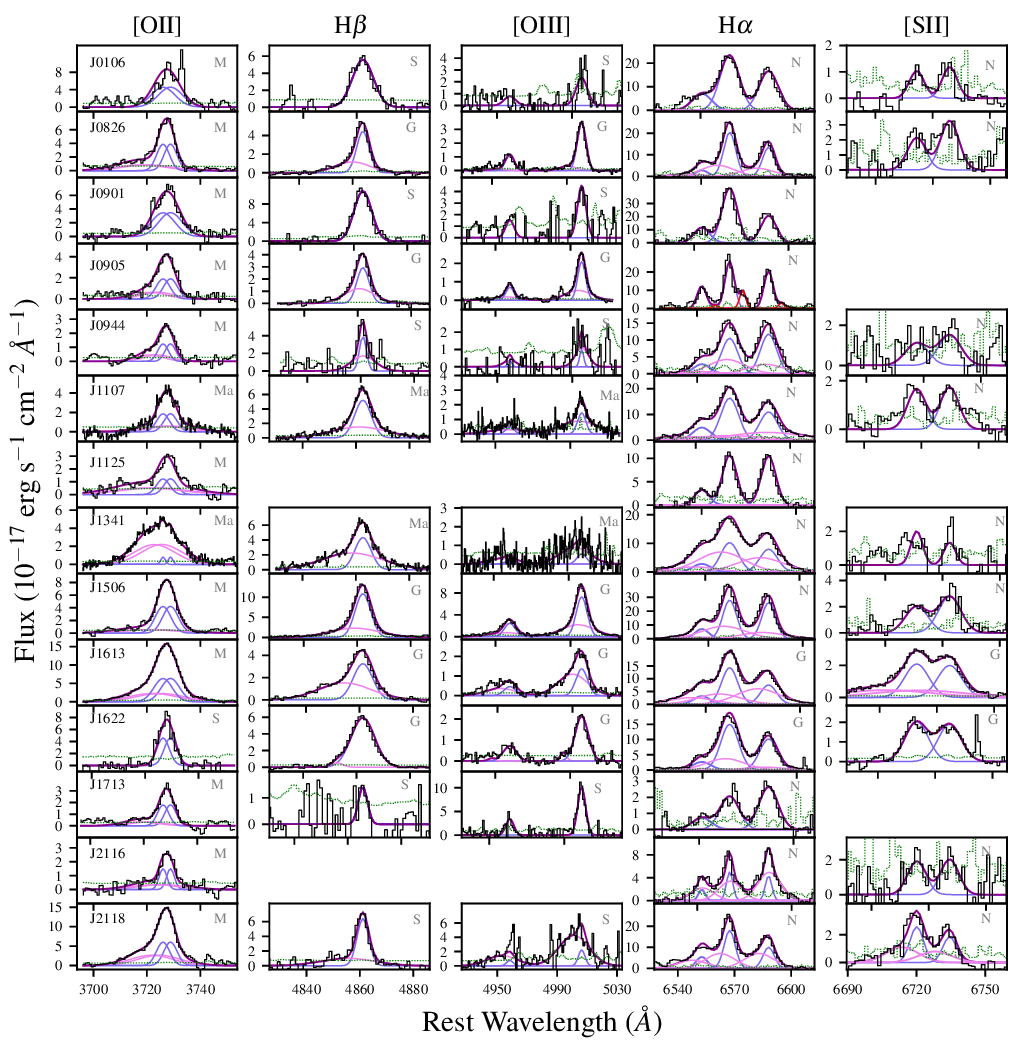}
  \caption{Fits to the nebular emission lines in the fourteen galaxies in our sample. Each row represents one object and each column from left to right is \ion{[O}{2]}$\lambda\lambda$3726,3729, H$\beta$, \ion{[O}{3]}$\lambda\lambda$4959,5007, the H$\alpha$+\ion{[N}{2]}$\lambda\lambda$6549,6585 blend, and \ion{[S}{2]}$\lambda\lambda$6717,6731. The grey letters represent the instrument used to obtain each spectrum: MMT (M), Magellan/MagE (Ma), Gemini/GMOS (G), Keck/NIRSPEC (N), or SDSS (S). The purple solid line shows the best fit to each emission line, the light blue and pink ones refer to the narrow and broad Gaussian components of the fit, respectively. We include a broad component when it improves the reduced $\chi^2$ of the fit significantly. The error spectrum is shown as a dotted green line. Spectra are omitted where the SNR is too low to identify any significant emission line. The identification of broad emission is indicative of outflowing material, and since the broad emission is seen in the forbidden lines, this suggests that the outflow originates from the ISM (rather than any hidden AGN broad-line region).}
  \label{fig:Fits}
\end{figure*}

The multicomponent fits to the nebular emission lines for the galaxies in our sample are shown in Fig.~\ref{fig:Fits}. The various emission lines are not fit simultaneously since the data sets have different resolutions and SNR. Moreover, the lines span a broad range in wavelength and extinction might impact them differently. The MMT/MagE data cover the \ion{[O}{2]} doublet spectral region. We assume the \ion{[O}{2]} doublet lines have identical kinematics (i.e., same velocity widths and shifts in the Gaussian fit components). We set the flux ratio \ion{[O}{2]}$\lambda$3729/\ion{[O}{2]}$\lambda$3726 to 1.005 as the spectra do not have sufficient resolution to fit them separately. We fix the \ion{[O}{2]} ratio to reflect the typical electron density of the ISM in our sources as estimated using the \ion{[S}{2]} emission lines (see Section~\ref{subsection:electron_density}; \citealp{san16}).
The \ion{[O}{2]} lines generally require two Gaussian components to fit their asymmetric profiles. The only exceptions are J0106, J0901, and J1622.


The H$\beta$ and \ion{[O}{3]} spectral region is covered by the SDSS data for 8/14 galaxies in our sample, and by the GMOS data for the remaining 6/14 galaxies (see Section \ref{subsection:gmos}). As in the case of the \ion{[O}{2]}, we adopt the same kinematics for the \ion{[O}{3]} doublet lines, and we fix their amplitude ratio \ion{[O}{3]}$\lambda$4959/\ion{[O}{3]}$\lambda$5007 to 0.337 to match the transition strengths \citep{sto00}. While we allow the H$\beta$ profile to have a different kinematic structure than that of \ion{[O}{3]}, we find consistent results between the line in terms of velocity widths and centroids of the narrow and broad components. The low SNR prevents us from performing a reliable fit of these lines for J1125 and J2116. Both H$\beta$ and \ion{[O}{3]} are well described by one Gaussian in 3 galaxies (J0106, J0901 and J1713), and by two Gaussians in the remaining 9 galaxies. 

Finally, we use the NIRSPEC data to fit the H$\alpha$, \ion{[N}{2]} and \ion{[S}{2]} emission lines for 12/14 galaxies in our sample, and the GMOS data for J1613 and J1622. All the emission lines in this spectral region are forced to have the same kinematics (velocity offsets and widths), while the amplitude of each component is allowed to vary independently. This choice is justified by the complex emission line profiles of H$\alpha$ and \ion{[N}{2]} that blend together, and by the low SNR of the \ion{[S}{2]} lines of the spectra in our sample. We do not fix the \ion{[N}{2]} doublet flux ratio to be 1:3, as the \ion{[N}{2]} $\lambda$6549 line for some of our galaxies falls at the edge of the NIRSPEC bandpass, where the spectra have higher fluxing errors. However, we find the \ion{[N}{2]} doublet flux ratio to be very close to the theoretical value in most cases, with a mean value of 0.38. We also perform fits fixing the \ion{[N}{2]} doublet ratio to 1:3 and find that the kinematics and fluxes of the H$\alpha$ and \ion{[N}{2]} emission lines change by $<$10\%. The broad \ion{[N}{2]} doublet ratio is set to be the same as the narrow \ion{[N}{2]} doublet ratio. 
The ratio of the density-sensitive \ion{[S}{2]} doublet is allowed to vary, but it is restricted to be within 20\% of the range of permitted values  \citep[0.43$-$1.5;][]{tay10, men14}. The H$\alpha$ and \ion{[N}{2]} kinematics are well parameterized by a single Gaussian in 5/14 galaxies (J0106, J0901, J0905, J1125 and J1713), and by two Gaussian components in the remaining 9/14 galaxies. Although we force \ion{[S}{2]} to have the same kinematics as H$\alpha$ and \ion{[N}{2]}, we are not able to fit a broad \ion{[S}{2]} component in any of the galaxies where it would be expected (from H$\alpha$) due to the low SNR, except for J1613 and J2118. Moreover, the low SNR prevent us from performing a reliable fit of the \ion{[S}{2]} doublet in four galaxies in our sample (J0901, J0905, J1125, and J1713). We also perform a fit of the \ion{[S}{2]} doublet lines not constrained by the H$\alpha$ and \ion{[N}{2]} kinematics. We obtain similar results but with larger uncertainties due to a larger number of free parameters.

Three of the galaxies have slight modifications to the fitting procedure: 1) J0905 is an unusual source that shows narrow redshifted H$\alpha$ $+$ \ion{[N}{2]} components; these offset features are fit separately using narrow Gaussian profiles with the same kinematics and are excluded from further analysis (marked in red in Fig.~\ref{fig:Fits}), and 2) the \ion{[O}{3]} kinematics for J0944 and J2118 are tied to the H$\beta$ kinematics due to the low SNR around the doublet emission lines. 

We correct all the emission line fluxes for dust extinction by comparing the Balmer decrement (H$\alpha$/H$\beta$) with the expected Case B value of 2.86 \citep{ost89}. Galaxies with Balmer decrements $<$ 2.86 (but consistent with 2.86 within the uncertainties) are assumed to have zero extinction. We adopt the Galactic extinction curve from \citet{car89} for galaxies with H$\alpha$/H$\beta$ $\geq$ 2.86, the interquartile range for extinction in our sample is E(B-V) = 0.18$-$0.70, with a median value of 0.36. 

Table \ref{table2} lists the full widths at
half-maximum (FWHM) corrected for instrumental resolution of both the narrow and broad Gaussian components of our spectral fits for the H$\alpha$ and \ion{[O}{2]} emission lines. We also report the velocity offset (v$_{off}$) of the broad component centroids with respect to the systemic redshift. The 1$\sigma$ errors on all measurements account for uncertainties in the fit parameters as well as covariance between parameters.


\section{Results}
\label{section:Results}

The following sections collect the results of this work. The main goal is to characterize the physical conditions of the starburst at the center of the galaxies in our sample that is driving powerful outflows.  We first investigate the kinematics of a suite of emission and absorption lines probing different scales of the same ionized outflowing gas. Then, we exploit an ensemble of emission line ratio diagnostics to derive quantities that regulate the emission of the \ion{H}{2} regions like electron density, metallicity, and ionization parameter. Lastly, we compare our findings with those of relevant comparison samples.

\subsection{Kinematics}
\label{subsection:kinematics}
The high SNR of the spectra employed in this study provides the unique opportunity of being able to measure the kinematics of \ion{[O}{2]}, \ion{[O}{3]}, H$\beta$ and H$\alpha$ emission lines independently. In Fig.~\ref{fig:Fits} we present the various observed emission lines and best fit line results for the fourteen galaxies in our sample. 
Although the nebular emission lines are fit separately, their line profile decompositions in narrow and broad components agree in 10/14 galaxies. Two of the remaining cases (J1125 and J1713) have the lowest SNR spectra covering \ion{[O}{3]}, H$\beta$ and H$\alpha$ in our sample. Both galaxies have \ion{[O}{2]} that clearly exhibits a broad and asymmetrical line profile. However, we do not include a broad component to other emission lines observed in these sources because the reduced $\chi^2$ of their fits do not improve significantly.
In the case of J1622, the \ion{[O}{2]} kinematics are well described by narrow lines only, while the \ion{[O}{3]}, H$\beta$ and H$\alpha$ fits require a broad component. Lastly, in J0905 we fit H$\alpha$ using a single narrow Gaussian, while \ion{[O}{2]}, \ion{[O}{3]} and H$\beta$ need an additional broad line (we note, however, that H$\alpha$ appears to have a secondary component, which may potentially be part of a broad line). We note that in all cases where a broad component is required for the best fit, the centroid of the broad component is blueshifted relative to that of the narrow component. We quantify the nebular emission line kinematics measured from our spectral fits using the FWHM and v$_{off}$ of each component. In Table \ref{table2} we report these values for H$\alpha$ and \ion{[O}{2]} only, as \ion{[O}{3]} and H$\beta$ exhibit kinematics that are very similar to H$\alpha$ and/or \ion{[O}{2]}.  

\begin{figure*}[hbtp!]
 \centering
 \includegraphics[width=0.85\textwidth]{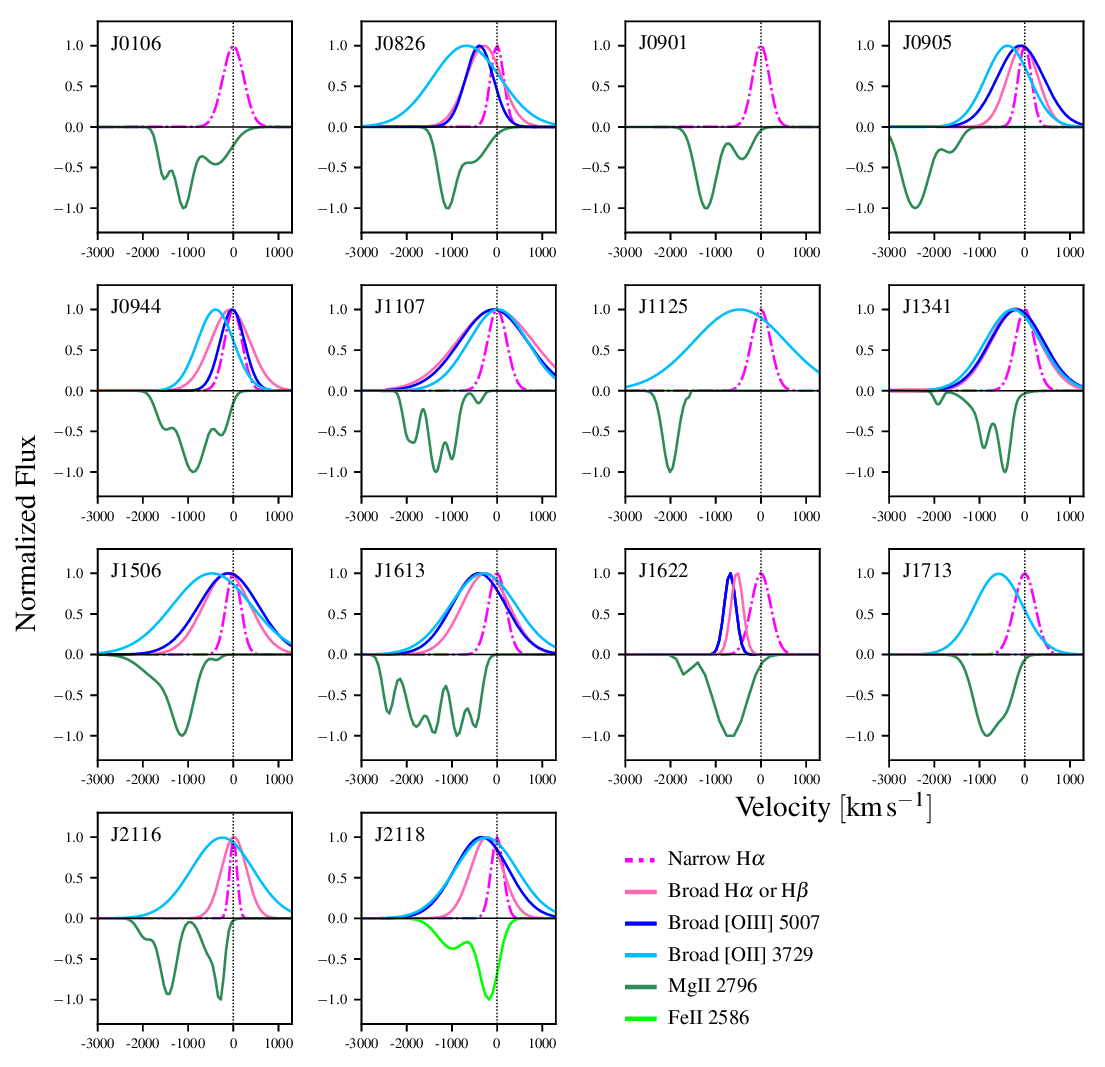}
 \caption{Comparison of velocity profile fits among selected emission and absorption lines for the galaxies in our sample. All profile fits are normalized to their emission or absorption flux peak, to facilitate comparison. The narrow H$\alpha$ emission line fit is displayed as a dot-dashed magenta line in each panel and represents the systemic redshift, in agreement with the redshift derived by the starlight (see Section~\ref{subsection:mmt}). Different outflowing gas tracers are shown as different color solid lines. Broad H$\beta$ is shown for J0905 and J1622. In J2118 \ion{Mg}{2} emission is observed, which obscures any underlying \ion{Mg}{2} $\lambda$2796 absorption feature, therefore we present \ion{Fe}{2} $\lambda$2586 instead for this galaxy, using KCWI data. The emission line velocity profiles show remarkable overall consistency, except for \ion{[O}{2]} $\lambda$3729 which tends to be more blueshifted compared to systemic in several sources. Emission and absorption lines probe different spatial scales of the same gas phase, and exhibit comparable maximum outflowing velocities in most of the galaxies in our sample.}
 \label{fig:Velocities}
\end{figure*}

Fig.~\ref{fig:Velocities} shows a comparison of the best spectral fits for a suite of emission and absorption lines for each galaxy in our sample. Each velocity profile is first normalized to its own emission or absorption line peak, to facilitate comparison. The narrow H$\alpha$ component is shown as a dot-dashed magenta line in each panel and traces the systemic redshift of the galaxy; the rest of the emission line components shown are broad.  We note that the broad \ion{[O}{2]} components (light blue solid line) are systematically wider than the H$\alpha$ broad components (pink solid line), with the exception of J0944 and J1107. The mean values of the broad FWHM for \ion{[O}{2]} and H$\alpha$ in our sample are 1573 and 1101 \kmps, respectively. Moreover, \ion{[O}{2]} shows larger blueshifts than H$\alpha$, except in source J1107. The mean values of v$_{off}$ for \ion{[O}{2]} and H$\alpha$ are 352 and 143 \kmps, respectively. 

Such line broadenings and blue velocity shifts clearly identify outflowing gas. We note that often the broad components contain some redshifted gas as well, compared to the narrow line profiles. 
The presence of a blueshift in the velocity centroid of the broad components is attributed to dust present in the host galaxy that obscures part of the redshifted outflows. We note that their SED fitting suggests a mean attenuation of A$\rm_V \sim 0.43$ (Tremonti et al., in prep.). We come back to this point in Section \ref{section:broad/narrow}.  

\begin{figure*}[hbtp!]
  \centering
   \includegraphics[width=0.7\textwidth]{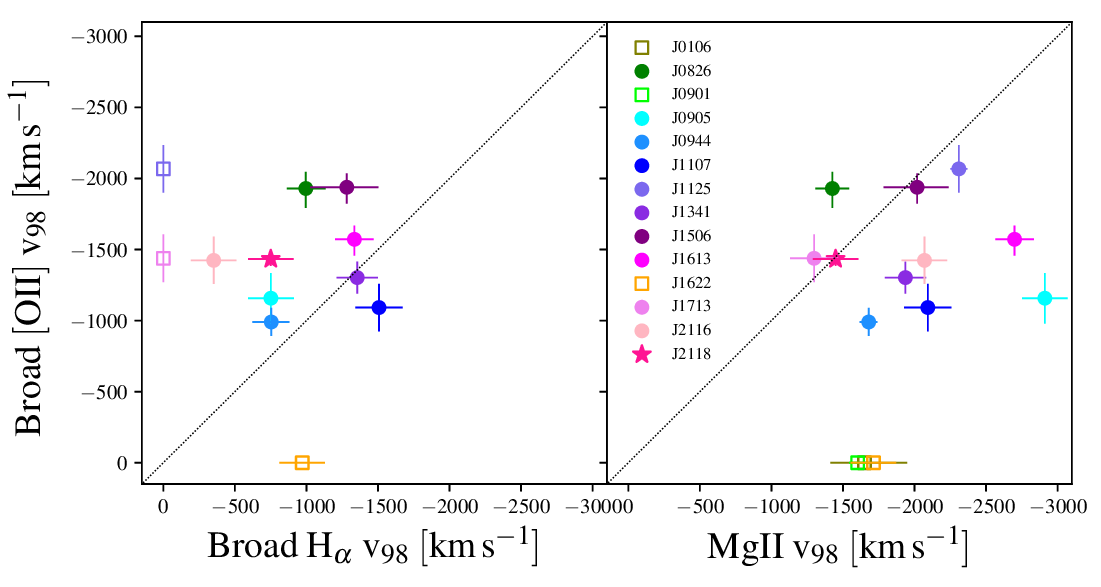}
  \caption{Broad \ion{[O}{2]} emission line kinematics compared to broad H$\alpha$ emission line (left), and \ion{Mg}{2} absorption line (right) ones as represented by the maximum measured velocity v$_{98}$. Errors on v$_{98}$ due to uncertainties in the fits are estimated varying the best-fit parameters in a range of $\pm$1$\sigma$ and measuring the resulting change in v$_{98}$. The dotted lines represent the 1 to 1 relation. The galaxies that have no broad \ion{[O}{2]} or H$\alpha$ emission lines detected are shown as empty squares. For J0905 v$_{98}$ is derived from the H$\beta$ broad emission line instead of the H$\alpha$. For J2118 v$_{98}$ is derived from the \ion{Fe}{2} $\lambda$2586 absorption line profile instead of the \ion{Mg}{2} $\lambda$2796, since \ion{Mg}{2} absorption is not detected for this galaxy. Most of the objects in our sample exhibit broad \ion{[O}{2]} maximum velocities comparable to those derived from the broad H$\alpha$ and \ion{Mg}{2} absorption lines.}
  \label{fig:comparison_vel}
\end{figure*}

The left panel of Fig.~\ref{fig:comparison_vel} compares the \ion{[O}{2]} and H$\alpha$ broad emission line kinematics as represented by v$_{98}$, which is an estimate of the maximum observed outflow speed (and is a lower limit to the actual maximum speed if the gas producing the blueshifted line wings is not moving directly towards the observer). 
The \ion{[O}{2]} maximum velocity is roughly 450 \kmps greater than that of H$\alpha$, although their kinematics are consistent for a few galaxies.

Fig.~\ref{fig:Velocities} compares the \ion{[O}{2]} emission line kinematics to fits of the  \ion{Mg}{2}$\lambda$2796 absorption lines for each galaxy.  \ion{Mg}{2} exhibits complex velocity profiles in our sources, with a mean value of v$_{98}$ of $-$1890 \kmps. Such large blueshifts clearly identify outflowing gas, observed in absorption. In the case of J2118 we do not detect \ion{Mg}{2} absorption and show the fit results to  \ion{Fe}{2}$\lambda$2586 instead. The lack of \ion{Mg}{2} absorption in this galaxy is most likely due to the detected \ion{Mg}{2} emission, which fills the underlying absorption trough. We note that 9/14 galaxies in our sample have less than 5\% of the \ion{Mg}{2} EW within 50 \kmps of the systemic redshift. While \ion{Mg}{2} emission line filling may be present for our sources it should not substantially affect our maximum velocity, as v$_{98}$ is typically far greater than the velocity of \ion{Mg}{2} when observed in emission. We will present results on \ion{Mg}{2} emission using high resolution spectra in an upcoming paper (Perrotta et al, in prep.).
We explore the possible reasons for the lack of \ion{Mg}{2} absorption near the systematic velocity below in Section~\ref{section:LyC_leak}. 

The various ions studied here probe the same cool gas phase (T $\sim$ 10$^4$ K). However, they could originate on different spatial scales and their physical properties could span a wide range of values. Most importantly, emission and absorption lines provide us different approaches to study outflowing gas. We return to this point in Section~\ref{section:broad/narrow}

\subsection{Electron Density}
\label{subsection:electron_density}

\begin{figure}[hbtp!]
  \centering
   \includegraphics[width=0.9\columnwidth]{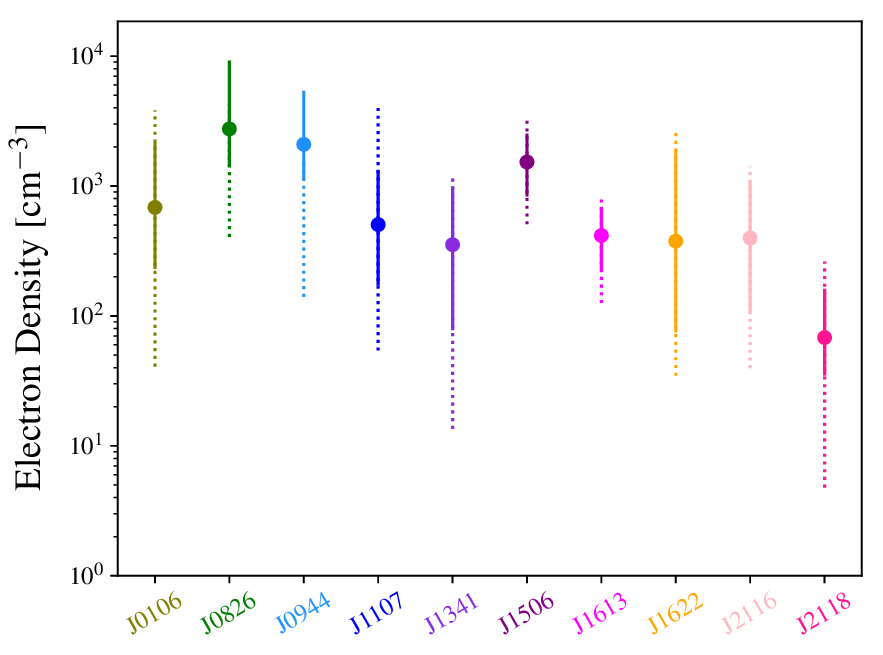}
  \caption{Electron densities calculated following the method described by Sanders et al. (2016) using narrow \ion{[S}{2]}$\lambda$6716/$\lambda$6731 doublet ratio. Errors on individual density measurements are estimated by converting the upper and lower 68th percentile uncertainties on the line ratio into electron densities. Solid error bars represent the errors derived using the uncertainties from the \ion{[S}{2]} constrained fit, and the dotted lines those from the \ion{[S}{2]} unconstrained fit.}
  \label{fig:elecdens}
\end{figure}

The electron density (n$_e$) of the ISM is one of the main physical quantities that govern the emission of \ion{H}{2} regions. The nebular emission-line ratios and derived quantities, such as the gas-phase metallicity and ionization parameter, probe the physical conditions in the central starburst and depend critically on measuring n$_e$. 

The electron density can be estimated from the ratio of the \ion{[S}{2]}$\lambda\lambda$6716,6731 doublet. The collisionally-excited forbidden lines are produced in low density gas, where the low number of collisions prevents the de-excitation of the excited state. Between the low density ($\lesssim$\,10\,cm$^{-3}$) and high density ($\gtrsim$\,10$^4$\,cm$^{-3}$) regimes this ratio provides a good measurement of the nebular gas density \citep[e.g.,][]{ost06}.

We employ the diagnostic relation from \citet{san16}, which assumes an electron temperature of T$_e$ = 10$^4$ K. For the two galaxies (J1613 and J2118) in our sample where the SNR is high enough to decompose the emission line profiles into separate narrow and broad components, we use the \ion{[S}{2]}$\lambda$6716/\ion{[S}{2]}$\lambda$6731 narrow line ratio. 
For the rest of the sample we use the \ion{[S}{2]}$\lambda$6716/\ion{[S}{2]}$\lambda$6731 total flux ratio. The results are shown in Fig.~\ref{fig:elecdens}. The errors on each density measurement are determined by converting the upper and lower 68th percentile uncertainties from the \ion{[S}{2]} constrained (solid line) and unconstrained (dotted line) fits on the line ratio into electron densities. The derived \ion{[S}{2]} doublet ratios range from 0.62 to 1.35, which correspond to an n$_e$ range from 68\,cm$^{-3}$ to 2750\,cm$^{-3}$. The median n$_e$ value across the full sample is 530\,cm$^{-3}$. This density range is substantially elevated with respect to typical \ion{H}{2} regions in the local universe, which generally have n$_e$ $\sim$ 50$-$100\,cm$^{-3}$ \citep[e.g.,][]{zar94}.

\begin{figure*}[t!]
  \centering
  \includegraphics[width=0.65\textwidth]{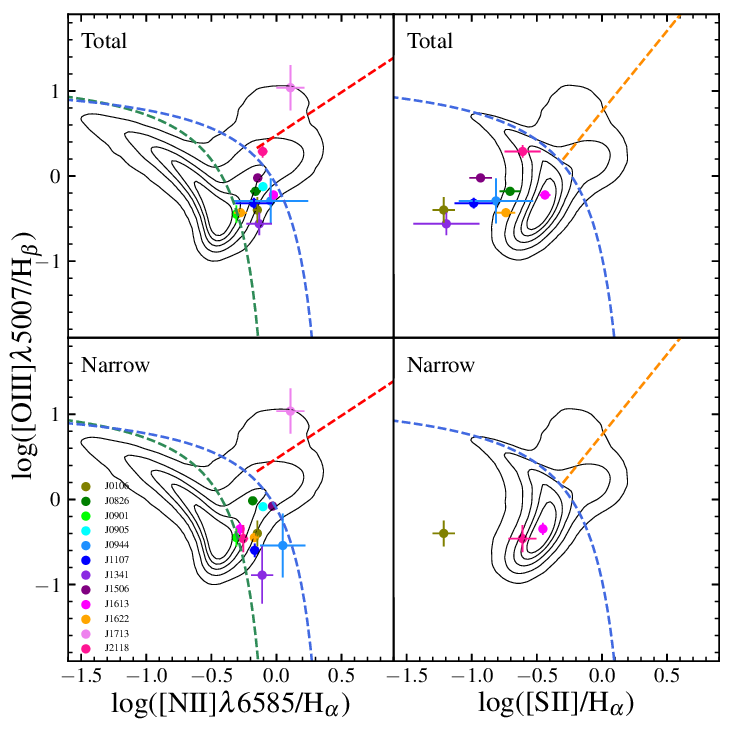}
  \caption{N2-BPT (left) and S2-BPT (right) diagrams for the total emission line flux (top panels) and the narrow component line flux (bottom panels) for the galaxies studied here. The green dashed lines delineate the empirical separation of star forming galaxies and AGN by \citet{kau03} in the N2-BPT plane. The blue dashed lines are theoretical curves derived by \citet{kew01} to show the location of maximal starburst galaxies in both diagrams.  Red and orange dashed lines from \citet{cid10} and \citet{kew06} separate LINER and Seyfert galaxies in the N2-BPT and S2-BPT planes, respectively. Contours show the location of SDSS DR8 galaxies for comparison (enclosing 30\%, 50\%, 70\%, 90\%, and 99\% of the galaxies). In the N2-BPT diagram our sample resides mainly in the composite region (with the exception of J1713, a type II AGN candidate), while in the S2-BPT diagram the total line fluxes in our sample are shifted to lower \ion{[S}{2]} to H$\alpha$ ratios than in SDSS galaxies.}
  \label{fig:BPT_TotNar}
\end{figure*}

The higher average electron densities we find in our galaxy sample are consistent with the characteristic electron densities observed in high redshift galaxies, which have values that are 5$-$10 times higher than  z\,$\sim$0 galaxies, with typical n$_e$ values of $\approx$ 200$-$400\,cm$^{-3}$ at z\,$\sim$2$-$3 \citep[e.g.,][]{mas14, ste14, san16, str17}. However, observations of some individual galaxies at z\,$\sim$2 suggest n$_e$ of $\sim$ 10$^3$\,cm$^{-3}$ \citep{hai09, leh09, qui09, bia10, shi14}. The high electron density implies the compact size of the \ion{H}{2} regions. If these high-z \ion{H}{2} regions follow the similar n$_e$-size relation found in the local galaxies
\citep{kim01}, their sizes should be less than 1 pc. We discuss how elevated n$_e$ values can affect the emission line production below in Section \ref{section:composite}.

\begin{figure*}[hbtp!]
  \centering
  \includegraphics[width=0.65\textwidth]{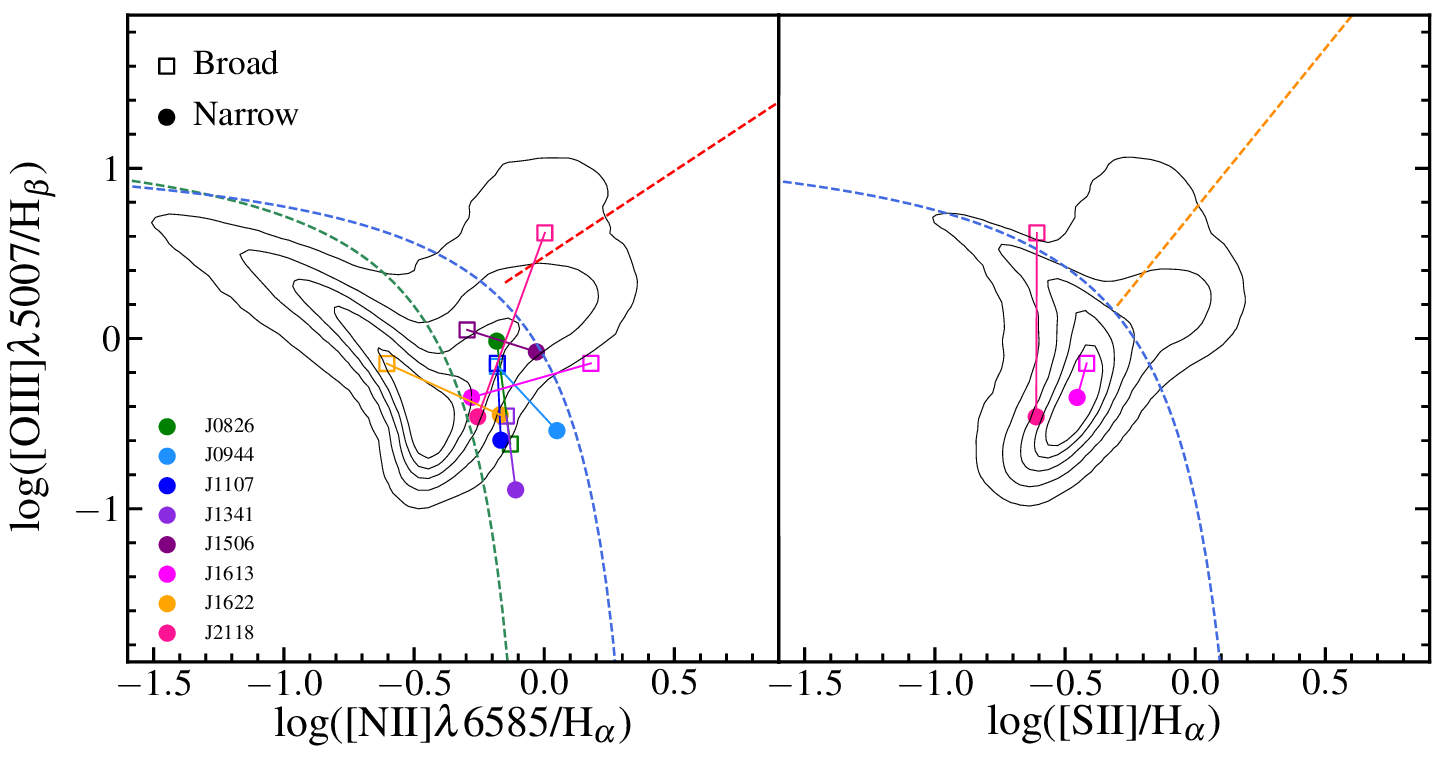}
  \caption{N2-BPT (left) and S2-BPT (right) diagrams comparing line ratios for the broad (open squares) and narrow (filled dots) emission line components for the galaxies in our sample. The two sources with SNR high enough to decompose the \ion{[S}{2]} emission line profile into separate narrow and broad components are shown in the S2-BPT plane. All dashed lines and contours are the same as in Fig.~\ref{fig:BPT_TotNar}. There is no obvious systematic variation of the \ion{[N}{2]} and \ion{[S}{2]} to H$\alpha$ ratios between the narrow and broad components, while the \ion{[O}{3]} to H$\beta$ ratio is routinely higher in the broad component than the narrow component in all but one galaxy in our sample.}
  \label{fig:BPT_Brd}
\end{figure*}

\subsection{BPT Diagnostic Diagrams}
\label{subsection:bpt}
Line ratios diagrams can be employed to distinguish between sources of ionizing radiation in emission line galaxies.  Following the work by \citet{bal81}, \citet{vei87} introduced the widely-used diagnostic diagrams commonly referred to as BPT diagrams. We consider the \ion{[O}{3]}$\lambda$5007/H$\beta$ vs. \ion{[N}{2]}$\lambda$6585/H$\alpha$ (N2-BPT), and \ion{[O}{3]}$\lambda$5007/H$\beta$ vs. \ion{[S}{2]}$\lambda\lambda$6717,6731/H$\alpha$ (S2-BPT) diagrams to characterize the galaxies in our sample.
 
Fig.~\ref{fig:BPT_TotNar} shows the N2- and S2-BPT diagrams, along with empirical and theoretical lines dividing galaxies excited by different mechanisms. Star forming galaxies occupy well defined regions in these diagrams. In particular, as metallicity increases, the sequence of star forming galaxies in the N2-BPT space elongates from high values of \ion{[O}{3]}$\lambda$5007/H$\beta$ and low \ion{[N}{2]}$\lambda$6585/H$\alpha$, and curves down to low \ion{[O}{3]}$\lambda$5007/H$\beta$ and high \ion{[N}{2]}$\lambda$6585/H$\alpha$. Moreover, galaxy stellar mass increases along this sequence, due to the correlation between stellar mass and gas-phase metallicity in star forming galaxies \citep{tre04}. The empirical lines dividing star-forming galaxies and AGN-hosted galaxies derived from SDSS are shown in Fig.~\ref{fig:BPT_TotNar} as green dashed lines \citep{kau03}, and the theoretical extreme starburst lines determined from photoionization and radiation transfer models are shown as blue dashed lines \citep{kew01}. The red and orange dashed lines represent the empirical lines separating LINER and Seyfert galaxies in the N2-BPT and S2-BPT planes, as derived by \citet{cid10} and \citet{kew06}. We assemble a comparison sample from the SDSS DR8, selecting galaxies within the redshift range 0.005\,$<$ z $<$\,0.1 to reduce aperture effects, and requiring 3$\sigma$ detection in the rest-frame optical emission lines featured in each diagnostic diagram. Emission line measurements and ancillary physical parameters are drawn from the MPA-JHU catalog for SDSS DR8\footnote{Available at \url{https://www.sdss.org/dr12/spectro/galaxy_mpajhu/}}. The grey contours enclose the 30\%, 50\%, 70\%, 90\% and 99\% of SDSS galaxies.

Fig.~\ref{fig:BPT_TotNar} shows the locations of our galaxies in the N2- (left) and S2-BPT (right) diagrams, where the top row uses line ratios determined from the total line flux, and the bottom row shows line ratios determined from the narrow line components only. 

The galaxies in our sample fall in or near the ``composite" region in the N2-BPT diagram, with the exception of J1713, which is a candidate type II AGN \citep{sel14}. Comparing the line ratios determined from the total line flux versus the narrow line flux, we find that there is not a bulk shift in the \ion{[N}{2]}$\lambda$6585/H$\alpha$ values, while the \ion{[O}{3]}$\lambda$5007 to H$\beta$ total flux ratio in all cases except one (J0826) is systematically higher than the corresponding narrow line flux ratio. 

We discuss in Section~\ref{section:AGN} possible AGN contribution to the line ratios.

Interestingly, most galaxies in our sample exhibit \ion{[S}{2]}$\lambda\lambda$6717,6731/H$\alpha$ values that are lower than   normal star forming galaxies, with 5/9 targets having lower total \ion{[S}{2]} to H$\alpha$ ratios than 99\% of SDSS galaxies. We discuss in Sections~\ref{section:composite} and \ref{section:LyC_leak} the possible causes of such low \ion{[S}{2]} to H$\alpha$ ratios.
The S2-BPT diagram for the narrow flux component (bottom right panel) includes the two galaxies (J1613 and J2118) with SNR high enough to decompose the \ion{[S}{2]} emission line profile in separate narrow and broad components. Both the total and narrow \ion{[S}{2]} to H$\alpha$ ratios of these two galaxies agree with those of normal star forming galaxies in the SDSS comparison sample. We also include J0106 as the  emission lines are fit with a narrow component only. The \ion{[S}{2]} to H$\alpha$ ratio for this galaxy is the lowest in our sample and is 0.37 dex lower than 99\% of the DR8 SDSS galaxies of comparable \ion{[O}{3]}/H$\beta$.

In Fig.~\ref{fig:BPT_Brd} we compare the locations of the line ratios for the narrow and broad components (filled dots and open squares, respectively) in the N2- (left) and S2-BPT (right) diagrams for the galaxies where we identify broad \ion{[O}{3]}, H$\beta$, H$\alpha$, \ion{[N}{2]}, and \ion{[S}{2]} lines. In the figure the flux ratios for the narrow and broad components in each galaxy are connected by a line, to ease comparison. The broad \ion{[O}{3]}$\lambda$5007/H$\beta$ ratio is routinely higher than the corresponding narrow line ratio, with the sole exception of J0826. We find that 5/8 galaxies have  \ion{[O}{3]}$\lambda$5007/H$\beta$ values for the broad component in the composite region of the N2-BPT diagram,  the ratios for J1613 and J2118 lie above the theoretical extreme starburst line \citep{kew01}, and the ratios for J1622 match those of normal star forming galaxies. The median \ion{[O}{3]} to H$\beta$ ratio for the narrow and broad components are 0.4 and 0.7, respectively. The systematic shift between the \ion{[N}{2]}$\lambda$6585 to H$\alpha$ ratios for the broad and narrow components in our sources is less clear. The median \ion{[N}{2]} to H$\alpha$ ratio for the narrow and broad components shift slightly higher from 0.67 to 0.69.

The \ion{[O}{3]}$\lambda$5007 to H$\beta$ ratio is sensitive to the hardness of the ionizing radiation field, and is useful to trace the ionization parameter of a galaxy \citep{bal81}. As shown in Section~\ref{subsection:kinematics}, the kinematics of the broad emission lines reflect that they probe outflowing gas. The higher ionization observed in the broad components could be caused by shocks associated with galactic outflows \citep{sha10}.  While the S2-BPT diagram can be used to identify shocks, unfortunately the low SNR of our spectra prevent us from exploring \ion{[S}{2]} broad lines in most of our sources.  The two galaxies where we can detect both broad and narrow \ion{[S}{2]}, J1613 and J2118, show similar \ion{[S}{2]}$\lambda\lambda$6717,6731/H$\alpha$ values for both components.

In this section we have shown that the galaxies in our sample fall in or very near the ``composite" region in the N2-BPT diagram, while exhibiting low \ion{[S}{2]} to H$\alpha$ ratios in the S2-BPT diagram. The position of a star forming galaxy on the BPT diagrams traces the ISM conditions and radiation field in the galaxy.  Several mechanisms can shift its location and mimic a composite star forming-AGN system: the raise of the hardness of the ionizing radiation field in a galaxy along the local abundance sequence or its electron density, the presence of shocks caused by galactic winds or mergers, the contamination of the line ratios by the diffuse ionized gas (DIG), complex geometrical gas distributions. As we will discuss in Section~\ref{section:Discussion}, the composite nature of the galaxies in our sample is more likely due to their extreme physical conditions than the presence of a buried AGN.

\begin{figure*}[t!]
  \centering
  \includegraphics[width=0.95\textwidth]{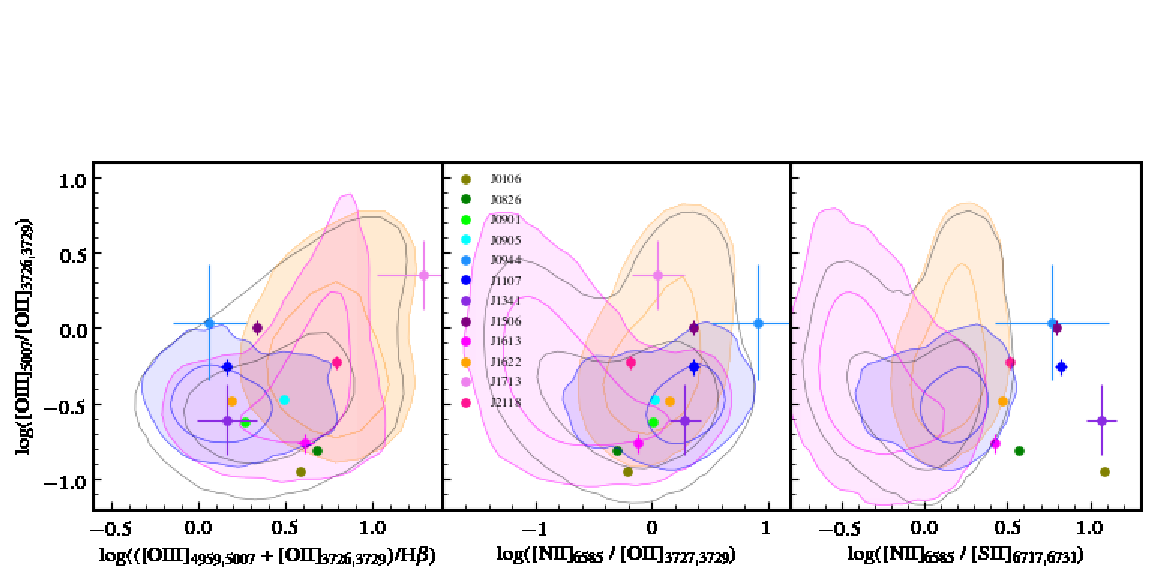}
  \caption{The ionization-sensitive ratio O32 (\ion{[O}{3]}$\lambda$5007/\ion{[O}{2]}$\lambda\lambda$3726,3729) plotted against abundance-sensitive diagnostics for our sample and the SDSS DR8 comparison sample. Light grey contours enclose the 80\%, and 99\% of the SDSS galaxies, while blue and magenta contours enclose the 80\% and 99\% of the high (M$_*$ $>$ 10$^{10.52}$ M$_{\odot}$) and low (M$_*$ $<$ 10$^{9.25}$ M$_{\odot}$) mass star-forming SDSS galaxies, respectively. The yellow contours illustrate the location of 80\%, and 99\% of the SDSS AGN-host galaxies. Left panel: R23 ratio ((\ion{[O}{3]}$\lambda\lambda$4959,5007 + \ion{[O}{2]}$\lambda\lambda$3726,3729)/H$\beta$; \citealp{pag79}). Central panel: N2O2 ratio (\ion{[N}{2]}$\lambda$6585/\ion{[O}{2]}$\lambda\lambda$3726,3729; \citealp{eva85, eva86, dop00}). Right panel: N2S2 ratio (\ion{[N}{2]}$\lambda$6585/\ion{[S}{2]}$\lambda\lambda$6717,6731; \citealp{dop13}).}
  \label{fig:O32_plot}
\end{figure*}

\subsection{Ionization and metallicity}
\label{subsection:ion_met}
Knowledge of the ionization parameter is crucial in understanding the properties of the ionizing sources as well as their impact on the surrounding ISM and outflowing gas. 
This parameter is typically measured using the ratio of two emission lines from the same atomic species that are in different ionization states. Fig.~\ref{fig:O32_plot} shows the commonly-employed ionization parameter diagnostic O32 (\ion{[O}{3]}$\lambda$5007/\ion{[O}{2]}$\lambda\lambda$3726,3729) plotted against abundance-sensitive ratios for the galaxies in our sample and in SDSS DR8 for comparison. 

The left panel shows O32 versus a widely-used optical metallicity diagnostic, the  R23 ratio ((\ion{[O}{3]}$\lambda\lambda$4959,5007 + \ion{[O}{2]}$\lambda\lambda$3726,3729)/H$\beta$; \citealp{pag79}). Our sample exhibits similar O32 and somewhat lower R23 ratios than SDSS galaxies, with median values of 0.3 and 2.5, respectively, compared to the full SDSS sample which has median values of 0.3 and 2.8.
The blue and magenta contours enclose the 80\% and 99\% of the high (M$_*$ $>$ 10$^{10.52}$ M$_{\odot}$) and low (M$_*$ $<$ 10$^{9.25}$ M$_{\odot}$) mass star-forming SDSS galaxies. They have median O32 values of 0.3 (high mass) and 0.4 (low mass), and average R23 values of 1.3 (high mass) and 4.6 (low mass). The composite SDSS galaxies occupy the region between these two in the O32-R23 space. The AGN-host galaxies (yellow contours, identified by the \citet{kew01} line), have average O32 and R23 values of 0.5 and 5.8, respectively. 

The galaxies in our sample exhibit ionization properties and R23 values consistent with those of the high mass tail of SDSS star-forming galaxies.
We note that J1713 is the only clear AGN candidate in our sample, and it lies in the AGN locus with high O32 and low R23.  


 R23 is sensitive to abundance but is double-valued as a function of metallicity. It increases with metallicity at low gas-phase O/H as the number of oxygen atoms increases, and it reaches a maximum at slightly less than solar abundance. Then R23 decreases again at high gas-phase O/H because the oxygen acts as an efficient cooler, reducing the gas temperature and consequently the number of collisionally-excited oxygen ions. Therefore, it is crucial to establish which solution branch applies when R23 values are low. The degeneracy can be resolved by the use of an additional parameter such as N2O2 (\ion{[N}{2]}$\lambda$6585/\ion{[O}{2]}$\lambda\lambda$3726,3729; \citealp{eva85, eva86, dop00}). N2O2 exhibits a remarkably tight correlation with metallicity above Z = 0.4Z$_{\odot}$, with an rms error of 0.04 \citep{kew02}. The reasons why N2O2 is highly sensitive to metallicity are twofold. First, nitrogen has a large secondary component of nucleosynthesis at high abundance, which causes an increase of N2O2, and second, the nebular electron temperature declines as the abundance increases. This leads to a strong decrease in the number of collisional excitations of the \ion{[O}{2]} lines relative to the lower energy \ion{[N}{2]} lines at high abundance. Moreover, N2O2 is almost independent of the ionization parameter because of the similar \ion{[N}{2]}$\lambda$6594 and \ion{[O}{2]}$\lambda$3726 ionization potentials, making this ratio the most reliable metallicity diagnostic in the optical. 

The central panel of Fig.~\ref{fig:O32_plot} shows O32 versus N2O2 for our galaxies and the SDSS comparison sample. Our galaxies exhibit high N2O2 ratios, with an average value of 1.3, in line with the most massive SDSS star-forming galaxies, suggesting high metallicities \citep{kew02, kew19}. This result implies that the R23 values in our sample are low because they are part of the high abundance solution branch. We apply a reddening correction to the \ion{[N}{2]} and \ion{[O}{2]} lines (see Section~\ref{sectio:Emission Line Fitting}), although our sample has uncertain dust content and geometry. 
While an accurate determination of the gas metallicity in our sample is beyond the purpose of this work, it is clear that our galaxies have high metallicities.

In the right panel of Fig.~\ref{fig:O32_plot} we show O32 versus N2S2 (\ion{[N}{2]}$\lambda$6585/\ion{[S}{2]}$\lambda\lambda$6717,6731; \citealp{dop13}) for our galaxies and the SDSS comparison sample. At high metallicity, nitrogen is a secondary nucleosynthesis element and sulphur is a primary $\alpha$-process element. They have similar excitation potentials, and in the high metallicity range their line ratio is a function of metallicity, due mainly to the different nucleogenic status of the two elements. The N2S2 diagnostic is not as useful as N2O2 for the determination of abundance because it is considerably more sensitive to the ionization parameter, but it has the strong advantage that reddening corrections are negligible.  Our sample exhibits high N2S2 ratios, with an average value of 5.3, again implying high metallicity \citep{kew02, kew19}. Some of the targets in our sample have N2S2 values similar to those of the most extreme high mass SDSS star-forming and AGN host galaxies. However, both these galaxy populations have average N2S2 of 1.5, more than three times lower than the average value for our sample.

Lastly, we note that two commonly-used metallicity calibrations by \citet{mcg91} and \citet{zar94a} infer derived log(O/H) + 12 = 9.0 and log(O/H) + 12 = 8.9, respectively, for galaxies in our sample. These values are in line with those inferred using the N2O2 and N2S2 diagnostics.

\begin{figure*}[hbtp]
  \centering
  \includegraphics[width=0.95\textwidth]{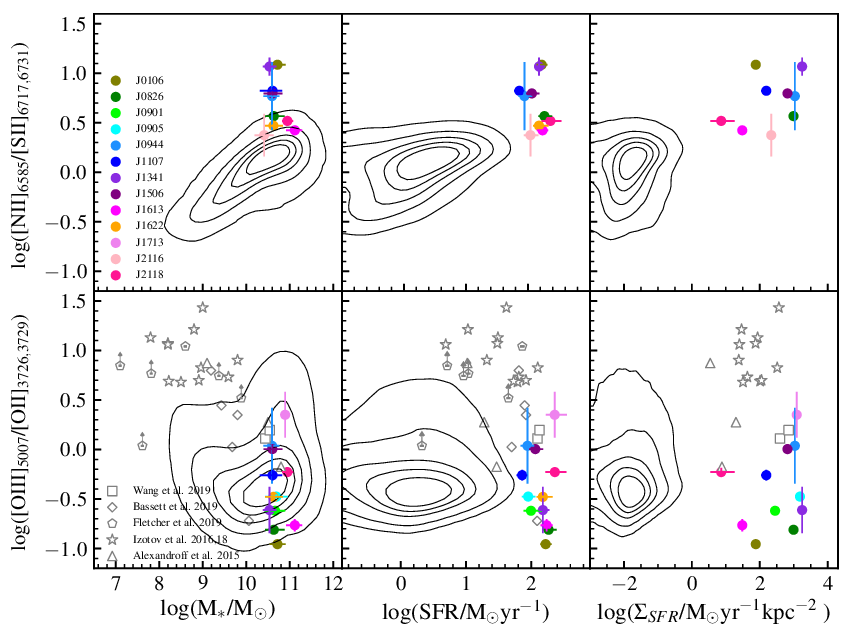}
  \caption{Top panels: total \ion{[N}{2]}$\lambda$6585 to \ion{[S}{2]}$\lambda\lambda$6717,6731 flux ratio compared to stellar mass (left), star formation rate (central), and star formation rate surface density (right). Bottom panels: total \ion{[O}{3]}$\lambda\lambda$5007 to \ion{[O}{2]}$\lambda\lambda$3726,3729 flux ratio compared to stellar mass (left), star formation rate (central), and star formation rate surface density (right). The grey contours represent SDSS DR8 data with contours at 25\%, 50\%, 75\%, 90\%, and 99\%. Black empty symbols are Lyman continuum leaking galaxies: z$\sim$0.3 \ion{[S}{2]}-weak galaxies (squares; \citealp{wan19}), low-redshift Green Pea galaxies (stars; \citealp{izo16b, izo16, izo18a, izo18b}), low-redshift Lyman Break Analogs (triangles; \citealp{ale15}, z$\sim$3 star-forming galaxies (diamonds; \citealp{bas19}), and z$>$3 LACES galaxies (pentagons; \citealp{fle19}). Five targets from \citet{fle19} are not detected in \ion{[O}{2]}, the O32 values are 3$\sigma$ lower limits.}
  \label{fig:gal_prop}
\end{figure*}

\subsection{Comparison with galaxy properties}
\label{subsection:connection}
In this section we investigate how the N2S2 and O32 line ratios depend on the physical properties of the galaxies studied in this paper, as compared to other galaxy populations.

In Fig.~\ref{fig:gal_prop} in the top row we show N2S2 versus the galaxy stellar mass (M$_*$), star formation rate (SFR), and star formation rate surface density ($\Sigma_{SFR}$) for galaxies in our sample as well as in SDSS.  We see in the upper left panel the well known relation between galaxy mass and metallicity (as seen in N2S2) in SDSS. 
The galaxies in our sample are uniform in M$_*$ with values comparable to the high mass tail of SDSS galaxies.
Our galaxies also have high N2S2, higher even than the typical N2S2 ratio at the high masses of our galaxies.  This likely reflects the {\it lack} of S2 in our sources, as seen in the S2-BPT diagram above.  In the middle and right panels it is clear that our galaxies have extremely high SFR and $\Sigma_{SFR}$ values, beyond SDSS galaxies.

In the lower panels we investigate the relationship between the O32 diagnostic and galaxy properties, again for galaxies in our sample and in SDSS.  We also show  known Lyman continuum (LyC) ``leakers" at low and high redshift \citep{ale15, izo16, izo16b, izo18a, izo18b, bas19, wan19, fle19}. As pointed out in Section~\ref{subsection:ion_met}, our sample shows O32 ratios comparable to the most massive SDSS galaxies, and N2S2 ratios similar to some of the most extreme SDSS galaxies. However, the implied average metallicity from N2S2 is much higher than that of the bulk of any SDSS galaxy population. As discussed in Section~\ref{section:LyC_leak}, LyC leakage may affect \ion{[N}{2]} and \ion{[S}{2]} differently, producing a deficiency of \ion{[S}{2]} and consequently, anomalously high N2S2 observed values.

An interesting comparison with our sample in the lower panels of Fig.~\ref{fig:gal_prop} is with confirmed LyC leakers, namely galaxies with an estimated fraction of ionizing, Lyman continuum photons ($\lambda$ $<$\,912\,\AA) that escape into the IGM that is greater than zero ($f_{esc}$(LyC)$>$\,0). 
Our sample exhibits some distinctive characteristics of known LyC leakers but differs in other crucial properties.
Most of the LyC leakers are substantially less massive than our galaxies. They span a wide range (3.7 dex) of M$_*$, with an average value of 10$^{9.1}$ M$_{\odot}$, $\sim$1.5 orders of magnitude lower than the average M$_*$ for our sample. LyC leakers display a broad range of O32 values (2.15 dex). Their average O32 is 1.2 dex higher than in our sample, however, the most massive LyC leakers overlap well with the O32 values of the compact starburst galaxies considered in this work. 
 The SFR and $\Sigma_{SFR}$ values of the LyC leakers are more similar to those of our galaxies. Specifically, in these samples LyC leakers have an average SFR of 37 M$_{\odot}$\,yr$^{-1}$ and an average $\Sigma_{SFR}$ of 147 M$_{\odot}$\,yr$^{-1}$\,kpc$^{-2}$; these values are four times lower than the average values in our sample. It is worth noting that both the LyC leakers and our sample are entirely distinct from the SDSS galaxy population in terms of having very high $\Sigma_{SFR}$ values.

While there are not N2S2 ratios reported for the LyC leakers presented in Fig.~\ref{fig:gal_prop}, some have metallicity estimates ranging from log(O/H) + 12 = 7.62 to log(O/H) + 12 = 8.16 \citep{izo16, izo16b, izo18a, izo18b}. These LyC leakers are considerably less metal-rich than our galaxies, as expected by their lower stellar masses. (Such low values correspond to a regime where N2S2 is not sensitive to metallicity, with values around 0.3 \citep{kew02, kew19}. The most massive LyC leakers shown in Fig.~\ref{fig:gal_prop} have derived metallicity in the range 8.18\,$<$ log(O/H) + 12 $<$\, 8.86 \citep{ale15, bas19, wan19}, where 8.7 corresponds to solar metallicity. (These values imply an N2S2 $<$\,3.2, \citealp{kew02, kew19}).
We discuss in Section~\ref{section:LyC_leak} below whether the galaxies in our sample are LyC leaker candidates.

\section{Discussion}
\label{section:Discussion}

We next discuss our results, including possible origins of the kinematically broad flux emission (Section~\ref{section:broad/narrow}). In Section~\ref{section:AGN} we examine the possible contribution of AGN to the observed emission lines  and then consider several additional mechanisms that can affect the location of our sample in the line ratio diagnostic plots (Section~\ref{section:composite}). We then review the properties of the galaxies in this study as potential LyC leaker candidates (Section~\ref{section:LyC_leak}).

\subsection{Interpreting Broad Emission Lines as Tracers of Galactic Outflows}
\label{section:broad/narrow}

Galactic winds are typically identified through their kinematic signatures. Winds seen in emission are detected as broad lines identified alongside a narrower component resulting from star forming regions in the galaxy \citep[e.g.][]{new12, fre19}. As shown in Section \ref{subsection:kinematics}, the emission lines in 12/14 galaxies in our sample require a broad$+$narrow Gaussian decomposition for at least one of the emission lines studied in this work (i.e. \ion{[O}{2]}$\lambda\lambda$3726,3729, H$\beta$, \ion{[O}{3]}$\lambda\lambda$4959,5007, H$\alpha$, \ion{[N}{2]}$\lambda\lambda$6549, 6585, and \ion{[S}{2]}$\lambda\lambda$6716,6731). The mean values of the velocity dispersion ($\sigma$) in the \ion{[O}{2]} and H$\alpha$ broad components in our sample are 670 and 470 \kmps, respectively. The broad components are also offset in their centroid velocities from the narrow components, blueshifted by mean values of 352 and 143 \kmps in \ion{[O}{2]} and H$\alpha$, respectively.  Such line broadening and blueshifts are interpreted in galactic spectra as outflowing gas. In many cases for the galaxies in our sample, the broad components exhibit some redshifted emission as well compared to the narrow line profiles, though the velocity centroids are always blueshifted. We attribute this to dust present in the host galaxy that obscures a portion of the redshifted outflows.

Star formation-driven outflows are observed in galaxies of all stellar masses, with an occurrence that correlates  with star formation properties, specifically SFR, the offset from the main sequence of star formation, and $\rm\Sigma_{SFR}$ \citep[e.g.,][]{kor12, rub14, hec15, chi15, for20}. 
Our sample probes high $\rm\Sigma_{SFR}$, and as expected it presents a high incidence of broad emission lines. However, many aspects are important in interpreting trends of outflow characteristics with galaxy properties. For example, the capability to detect an outflow depends on the strength of the wind signatures, along with the SNR and spectral resolution of the data. Slower or weaker winds are more difficult to identify, especially using nebular emission lines as the broad components can be difficult to separate from the narrow emission from star formation. Therefore, a note of caution is in order when using the incidence of broad lines as a function of galaxy properties. 
Also, differences in sample selection and assumptions made in the analysis may result in different conclusions. For example, there have been claims of a strong dependence of the outflow incidence on $\rm \Sigma_{SFR}$ in high-redshift star-forming galaxies, though the existence and location of a threshold in $\rm \Sigma_{SFR}$ is somewhat unclear \citep{new12, dav19}.
In a forthcoming paper (Davis et al. in prep) we investigate scaling relations between outflow and galaxy properties for 46 galaxies in our parent sample that we collected spectra for and review the biases related to the use of different outflow tracers.

\begin{figure*}[t!]
  \centering
   \includegraphics[width=0.7\textwidth]{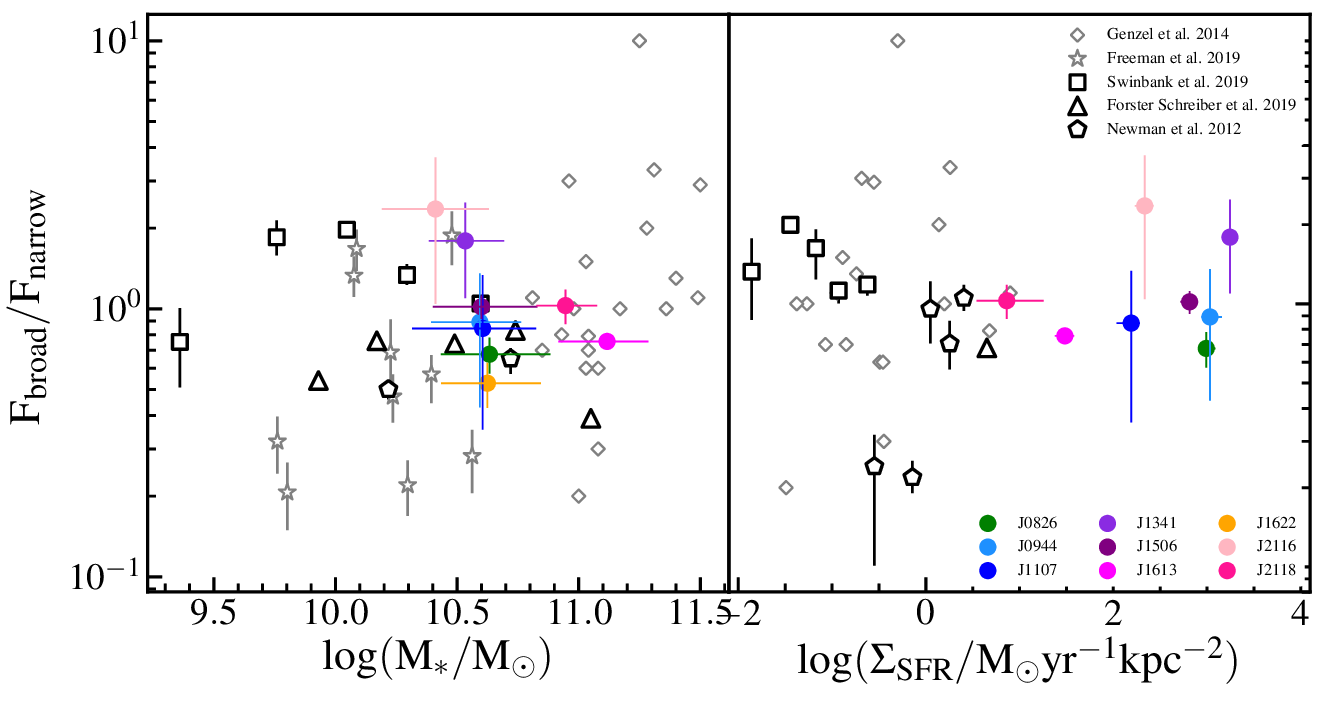}
  \caption{Broad-to-narrow H$\alpha$ flux ratio as a function of stellar mass (left panel), and star formation surface density (right panel) for our galaxies and some relevant star-forming galaxy samples. Stars are 10 star-forming galaxies at z $\sim$ 2 from the MOSDEF survey \citep{fre19}. Squares are the galaxy average values of 529 star-forming galaxies at z $\sim$ 1 from KMOS observations \citep{swi19}. Diamonds are 20 z = $1-2$ galaxies from \citet{gen14}. Pentagons are stacks of 27 z $\sim$ 2 star-forming galaxies from the SINS and zC-SINF surveys \citep{new12}. Triangles are stacks of 78 (left panel) and 33 (right panel) star-forming galaxies at 0.6 $<$ z $<$ 2.7 from the KMOS3D survey \citep{for19}. }
  \label{fig:BFR}
\end{figure*}

Emission and absorption lines provide us distinct approaches to investigate outflows. While emission lines derive from the projected signal of emitting gas filling the whole volume in front of and behind the galaxy, absorption lines probe only the gas along the line of sight illuminated by the central starburst. Furthermore, the absorption lines are sensitive to the density of the gas probed, while emission lines are sensitive to the density squared. This results in absorption lines providing access to lower density, weaker gas components. Comparing v$_{98}$ derived from the \ion{[O}{2]} emission lines and \ion{Mg}{2} absorption lines in the right panel of Fig.~\ref{fig:comparison_vel}, we see that generally the \ion{Mg}{2} maximum velocities are higher (though they are consistent with \ion{[O}{2]} emission for several galaxies). This might be explained if the outflowing gas has a lower density, on average, which makes it easier to accelerate. It is also reasonable that absorption line velocities may be higher than emission line velocities, on average, as emission lines can probe gas that is both in front of and behind the galaxy. This can produce a redshifted wing in  emission profiles that shifts both the central velocity and the velocity at which 98\% of the total EW is detected towards smaller values.

Broad emission lines have also been used to constrain outflow properties beyond kinematics. 
The broad to narrow flux ratio (BFR) of H$\alpha$ has been shown in the literature as a function of galaxy parameters and used to infer the mass loading factor ($\eta$ = outflow mass rate/SFR). Adopting a model that describes the outflow  geometry and physical conditions, it is possible to convert the observed H$\alpha$ BFR into an estimate of $\eta$ \citep[][]{ste10, gen11}. This approach has been used to identify a possible threshold in star formation properties above which a galaxy has the ability to power outflows \citep[e.g.,][]{new12, fre19}. In particular, the inferred $\eta$ has been found to strongly correlate with $\rm\Sigma_{SFR}$ within some galaxy samples. Therefore, a $\rm\Sigma_{SFR}$ threshold has been proposed that dictates when star formation feedback may break through the dense gas layers in the galactic disk and launch a large-scale outflow. 

For comparison to other studies, we parameterize the broad emission we measure using the BFR. Fig.~\ref{fig:BFR} shows the H$\alpha$ BFR as a function of M$_*$ and $\rm\Sigma_{SFR}$ for our sample and other relevant star-forming galaxies \citep{new12, gen14, swi19, for19, fre19}. Symbols with thick contours reflect stacked spectra, while grey symbols show results for individual galaxies.  
Fig.~\ref{fig:BFR} shows that when we consider samples spanning a wide dynamic range there is no correlation between BFR and M$_*$ or $\rm\Sigma_{SFR}$. 
Additionally, there is not clear evidence for a threshold in $\rm\Sigma_{SFR}$ above which outflows are launched.  Similarly, such a threshold is also not observed in low-redshift LIRG and ULIRG galaxies, even after correcting for the differential fraction of the gas content \citep{arr14}.

Trends of BFR with M$_*$ or $\rm\Sigma_{SFR}$ observed in previous studies are often in tension with theoretical expectations and numerical simulations \citep{new12, lil13, mur15, fre19, for20}. A reasonable explanation is that when  observations are used to infer global properties of outflows, the adopted assumptions regarding velocity, geometry, temperature, ionization source, and gas density are too simplistic and fail to capture the complexity of the outflows \citep{rup19}. Additionally, H$\alpha$ traces the warm ionized gas phase and much, if not most, of the outflowing mass is likely in an neutral atomic or molecular phase \citep{wal02, rup05, rup13, flu20, vei20}.
Given the potential systematic issues in detecting outflows using broad emission lines, a note of caution is warranted in interpreting any correlation between BFR and M$_*$ or $\rm\Sigma_{SFR}$, especially when different sample selections or analyses are involved.

\subsection{AGN Contamination}
\label{section:AGN}
All but one of the galaxies in our sample fall in the composite region in the N2-BPT 
diagram. Galaxies in this region are often interpreted as having contributions to their line ratios from both star formation and AGN, and it is therefore important to understand the possible AGN contribution in our sources. 

In general, we do not find evidence for widespread AGN activity in our sources. 
None of the galaxies in this study show evidence of an AGN  in their restframe near-ultraviolet and optical spectra (e.g. lack of very broad \ion{Mg}{2}, H$\beta$, or H$\alpha$). Additionally, none of the  sources would be classified as AGN based on their WISE mid-IR colors (the median W1$-$W2 of our sample is 0.35; \citealp{pet20}). They also do not satisfy the W1$-$W2 $>$ 0.8 (Vega) criterion of \citet{ste12} or the color-magnitude cuts of \citet{ass13} that include fainter sources. 

Ten galaxies in our sample (J0106, J0826, J0905, J0944, J1107, J1125, J1341, J1613, J2116, and J2118) have VLA 1.5 GHz continuum observations that allow us to place constraints on the ongoing radio AGN activity in these systems. The derived radio luminosities (L$\rm_{1.5GHz}$) span a $5.2 - 505\times$10$^{22}$ W\,Hz$^{-1}$, with a median value of 50$\times$10$^{22}$ W\,Hz$^{-1}$ \citep{pet20}. These L$\rm_{1.5GHz}$ are 3$\sigma$ below the radio excess threshold used by \citet{smo17} to identify AGN-dominated radio sources, and are compatible with being powered by the central starburst.  

Six galaxies in our sample were part of a $Chandra$ observing program targeting the 12 galaxies in the parent sample with the strongest indication for possible ongoing AGN activity based on emission-line properties \citep{sel14}. Three of the galaxies in this study (J1506, J1613, and J2118) have weak detections (4 X-ray counts each), implying an X-ray luminosity of L$\rm_x \approx$ 10$^{42}$ erg\,s$^{-1}$. The remaining three (J0826, J0944, and J1713) have upper limits corresponding to L$\rm_x <$ 10$^{43}$ erg\,s$^{-1}$. The derived X-ray luminosities are consistent with the known IR-based SFRs of these sources \citep{asm11, min14, sel14}.   

\citet{sel14}  classified J1713 as the most likely galaxy in their sample to host a type II AGN based on pseudo-BPT diagrams (e.g. \ion{[O}{3]}/H$\beta$ vs \ion{[O}{2]}/H$\beta$), and estimated a bolometric Eddington fraction of L$\rm_{bol}$/L$
\rm_{Edd}$ $\approx 0.02-0.13$. The new spectroscopic data and resulting line ratios for this galaxy lead to the same conclusion (see Fig~\ref{fig:BPT_TotNar}), as this galaxy does not lie in the composite region but is clearly in the AGN region of the BPT diagrams.  Moreover, J1713 is distinct from the rest of our sample in the ionization and abundance diagnostics plots (Fig.~\ref{fig:O32_plot}) and overlaps the SDSS AGN locus in these spaces.  We therefore conclude that this source does contain an AGN.  

J1506 exhibits a clear ($\sim$10$\sigma$) [NeV]3426\AA\, detection; this ion has a high ionization potential and is commonly used to trace AGN activity \citep[e.g.,][]{gil10}. \citet{sel14} estimate a ratio of the X-ray to [NeV] luminosity L$_x$/L$\rm_{[NeV]}$ = 4.9, implying a Compton-thick AGN (N$\rm_H > 10^{24}$ cm$^{-2}$). Under the assumption of the emission line being produced by an obscured AGN, \citet{sel14} find that the AGN would contribute $\sim$10\% of the mid-IR luminosity. However, [NeV] can also be powered by a very young (less than a few Myr) stellar population containing Wolf-Rayet and O stars \citep{abe08}. J1506 has a very young ($\sim$ 3 Myr) stellar population and the highest $\Sigma_{SFR}$ in our sample. Therefore, the observed [NeV] could be produced by the extreme conditions of the central starburst \citep{sel14}. [NeV]3426\AA\, emission is also detected in the outflowing component of another of our sources, J2118 \citep{rup19}. The derived luminosity L$\rm_{[NeV]}$=3.6$(\pm)1\times$10$^{40}$ erg\,s$^{-1}$, is three times lower than the averge for typical [NeV] emitters at similar redshift \citep{ver18} and could be produced by fast shocks with velocities of at least 300$-$400 \kmps \citep{bes00, all08}.

In summary, most of the galaxies in this study show no evidence for AGN activity based on X-ray and radio observations, optical emission lines, and infrared  colors. For the galaxies that may contain a dust-obscured accreting SMBH, the AGN contributes a small fraction of the bolometric  luminosity. While we cannot rule out past heightened AGN activity, multi-wavelength data for all of but one of these galaxies can be explained by their known star formation properties and the possible presence of shocks.

\subsection{Interpreting the BPT diagrams}
\label{section:composite}
In order to interpret the position of a galaxy in the N2- and S2-BPT diagrams and understand the gas ionization source(s), it is key to consider the mechanisms that can affect the integrated galaxy line ratios. 
In addition to the potential contribution from AGN as discussed above, here we consider the possible contributions from diffuse ionized gas (DIG) and shocks. 

Studies based on narrowband H$\alpha$ imaging have revealed that DIG can contribute substantially to the optical line emission in local galaxies \citep{zur00, oey07}. Typically, DIG exhibits enhanced forbidden-to-Balmer line ratios (e.g., \ion{[S}{2]}$\lambda\lambda$6717,6731/H$\alpha$, \ion{[N}{2]}$\lambda$6585/H$\alpha$, \ion{[O}{2]}$\lambda$3726/H$\beta$; \citealp{hoo03, mad06, vog06}) relative to \ion{H}{2} regions. Therefore, DIG contamination can move the location of a galaxy in the BPT diagrams towards composite or LINER-like regions \citep{sar06, yua10, keh12, sin13, gom16, bel16, bel16b}. \citet{zha17} and \citet{san17} have shown that DIG deviates from \ion{H}{2} regions more in emission-line diagrams featuring \ion{[S}{2]} or \ion{[O}{2]}, rather than \ion{[N}{2]}, and that DIG is characterized by a lower ionization parameter than \ion{H}{2} regions. Additionally, the fractional contribution of DIG emission to the Balmer lines ($f\rm_{DIG}$) is found to decline with increasing $\Sigma_{SFR}$ \citep{oey07, mas16, sha19}. Indeed DIG emission is negligible in typical high-redshift galaxies that are more highly star-forming \citep{whi14} and more compact \citep{vanderwel2014}. A substantial DIG contribution to the emission line ratios in our sample is in contrast to the low \ion{[S}{2]}/H$\alpha$ (Fig.~\ref{fig:BPT_TotNar})
observed values. Most importantly, similarly to high redshift galaxies, our sample is characterized by extremely high $\Sigma_{SFR}$ with a average value of 620 M$_{\odot}$\,yr$^{-1}$\,kpc$^{-2}$, roughly 4 order of magnitudes higher than the median SDSS $\Sigma_{SFR}$.  We therefore can safely assume negligible contamination from DIG ($f\rm_{DIG} \sim$ 0) when interpreting the BPT diagram locations of our galaxies. 

As discussed above, the presence of an AGN can also affect the location of galaxies in the BPT diagrams. 
As the contribution from an AGN increases, its host galaxy may migrate from the empirical sequence of \ion{H}{2} region emission toward the AGN portion of the diagnostic diagrams as a consequence of the increasing contribution from a harder ionizing radiation \citep{yua10}. However, in starburst+LINER systems the nature of the observed composite activity may be the result of non-AGN sources. In ultraluminous infrared galaxies (ULIRGs), extended LINER emission has been observed due to starburst wind-driven and merger-driven shocks \citep{sha10, ric10, ric11, sot12, ric15}. Moreover, shocks, common in ongoing mergers, can significantly enhance \ion{[O}{2]} relative to \ion{[O}{3]}, thus reducing the observed O32 which is used to probe the ionization state of a galxy \citep{ric15, epi18, bas19}.
Gas outflows and mergers can produce widespread shocks throughout a galaxy, which can substantially impact its emission line spectrum at both kpc and sub-kpc scales \citep{med15}. \citet{ric14} compared spatially resolved spectroscopy of 27 local ULIRGs to the spectra extracted from their brightest optical nuclear regions. Interestingly, they found that 75\% of the galaxies in their sample that would be classified as composite based on optical nuclear line ratios result from a sizable contribution from shocks to their emission line spectra. Therefore, shock emission combined with star formation can mimic “composite” optical spectra in the absence of AGN contribution.

Shock excitation can affect both low and high ionization line ratios. In slow shocks (v $<$ 200 \kmps) the shock front moves faster than the photoionization front caused by the shocked gas. This type of shock produces relatively weak high ionization lines, but strong low ionization species such as \ion{[S}{2]} and \ion{[N}{2]} \citep{ric11, ric15}.
In fast shocks (v $>$ 200 \kmps), the extreme ultraviolet and soft X-ray photons generated by the cooling of the hot gas behind the shock front produce a supersonic photoionization front that moves ahead of the shock front and preionizes the gas. This photoionization front is referred to as the photoionizing precursor, and it produces strong high ionization lines, while the hard radiation field from the shock front itself produces an extended partially ionized zone where low ionization lines such as [OI], [NI], and \ion{[S}{2]} are observed \citep{all08, kew19}. \citet{kew13} showed how local galaxies containing emission from either slow or fast shocks can result in composite locations in the BPT diagrams.

While the total luminosity of a shock depends only on its velocity and the gas density, the emission line spectrum depends strongly on the physical and ionization structure of the shock. This is determined primarily by the shock velocity, the magnetic parameter, and the metallicity. Moreover, the density may play a crucial role when it is sufficiently high for collisional de-excitation of forbidden lines to become important. The magnitude and direction of the emission line ratios shifts due to shocks are complex and difficult to predict.  
However, shocked emission tends to have higher \ion{[N}{2]}/H$\alpha$  and \ion{[S}{2]}/H$\alpha$ ratios compared to photoionized \ion{H}{2} regions \citep{all08, ric11}. 

Slow shock models \citep{ric11} can not simultaneously reproduce the (total and narrow) line ratios in the N2- and S2-BPT diagrams for our galaxy sample (Fig~\ref{fig:BPT_TotNar}). Fast shock $+$ precursor model grids from \citet{all08} produce too high \ion{[N}{2]} and \ion{[S}{2]} to H$\alpha$ ratios at given \ion{[O}{3]}/H$\beta$ compared to  the values for our sample. The \ion{[N}{2]}/H$\alpha$ and some of the \ion{[S}{2]}/H$\alpha$ observed ratios can be reproduced by some models that include only emission from the post-shock region. In particular, model grids that simultaneously match 70\% of our sample in the N2- and S2-BPT diagrams have high pre-shock density of $\sim1,000 ~\mathrm{cm}^{-3}$, solar or super solar metallicity, a magnetic field strength of $B < 32 \mu \mathrm{G}$, and a wide range of shock velocity values spanning $200-700~\kmps$. However, such a high pre-shock density would imply a post-shock density of $10,000 ~\mathrm{cm}^{-3}$ \citep{dop95}, which is an extreme and unlikely condition.  

The broad emission line ratios (Fig.~\ref{fig:BPT_Brd}) in both BPT diagrams reside within a wide range of model grids that include emission from either the post-shock region or both post- and pre-shock regions, with shock velocities of $200-1000 ~\kmps$ and pre-shock densities of $0.01-1,000 ~\mathrm{cm}^{-3}$. 
Our spectra do not have sufficient SNR to study the broad \ion{[S}{2]} component for most of the galaxies in our sample. However, in the two objects where we can identify a broad \ion{[S}{2]} line the line ratios are consistent with having a shock contribution, combined with star formation. 

It is extremely challenging to investigate the ionization source(s) in a galaxy when only a spatially-integrated spectrum is available, and it is
possible that the line ratios have contributions from multiple sources. However, shocks rarely dominate the global emission of a galaxy, and if a galaxy does contains shocks, there may also be contributions from star formation and/or an AGN. \citet{ric11} found that the enhanced optical line ratios from shocks are washed out by star formation and are thus easier to observe on the outskirts of galaxies where the level of star formation is lower. In our sample, where the optical light is dominated by young stars that formed within the central $\sim$ few hundred parsecs during recent starburst events \citep{dia21}, it is plausible that any signatures of potential shocks are washed out by the intense star formation present.

\subsection{LyC leakers candidates?}
\label{section:LyC_leak}

Next we discuss the possibility of the galaxies in our sample being Lyman continuum (LyC) leakers. This is of interest as it is currently unclear what sources are responsible for creating the epoch of reionization, which marks a crucial transition phase in the early Universe in which hydrogen in the IGM is transformed from a neutral to an ionized state 
\citep{fan06, kom11, zah12, bec15, boe19}.
Deep HST near-IR imaging indicates that primordial star-forming galaxies are capable of producing the bulk of the LyC photons needed to drive reionization \citep[e.g.,][]{bou12, oes13, rob15}.  It has been estimated that the escape fraction of LyC (f$_{esc}$(LyC)), i.e. the fraction of ionizing radiation released into the IGM, should be at least  $10-20$ percent on average \citep[e.g.,][]{ouc09, rob13, kha16}.

The increasing IGM neutral fraction at $\rm z > 5$ prevents a direct measurement of the LyC escaped from galaxies. Therefore, searches for LyC leakers are carried out at lower redshift to identify the indirect signs of LyC escape.  Many groups have observed such galaxies from low redshift up to z $\sim$ 4. Most of the confirmed LyC leakers have f$_{esc}$(LyC) below 0.15 \citep{lei13, bor14, izo16, lei16}. There are examples with estimated f$_{esc}$(LyC) as high as $0.45-0.73$ \citep{van16, deb16, sha16, bia17, van17, fle19, izo18a, izo18b}, however these galaxies are remarkably rare.

Some of the distinct observational signatures shared by the LyC leakers are strong Ly$\alpha$ emission with a double-peaked Ly$\alpha$ line profile \citep{ver15, ver17, van20}, high $\rm\Sigma_{SFR}$, high specific star formation rate (sSFR), and high ionization parameter traced by the O32 ratio \citep{izo18b, deb16, van20, cen20}. Some of them also show high n$_e$ \citep[e.g.][]{gus20}.

As seen in Section~\ref{subsection:connection}, the galaxies in our sample show some of the features common to known LyC leakers. In particular, both populations are substantially distinct from the SDSS galaxy locus in terms of having high $\rm \Sigma_{SFR}$; a physically-motivated model relating f$_{esc}$(LyC) to $\rm\Sigma_{SFR}$ was recently proposed \citep{sharma16,nai20, cen20}. 
The average O32 of known LyC leakers is around 1.2 dex higher than in our sample, however our sample overlaps well with the range of O32 values shown by the most massive LyC leakers. 
High O32 was initially used as a primary selection criterion to identify LyC leaker candidates, however, it was revealed that f$_{esc}$(LyC) does not correlate strongly with O32 (see \citealp{izo18b, nai18, bas19, nak20}, and discussions therein).

Most recently, \ion{[S}{2]} deficiency has been used as an empirical signpost to identify LyC emitter candidates \citep{wan19, ram20}. The \ion{[S}{2]} deficiency is a tracer of gas that is optically thin to ionizing radiation, allowing the escape of LyC photons. In a classical ionization-bounded \ion{H}{2} region, the \ion{[S}{2]} lines are produced in the warm partially ionized region near and just beyond the outer edge of the Str\"{o}mgren sphere. In a density-bounded nebula, the flux of ionizing photons from the central source is so large that the gas between the source and the observer is fully ionized. As a result, the ionizing radiation can escape because there is little or no neutral gas between the source and observer to absorb these photons. In this model, the outer partially-ionized \ion{[S}{2]} zone is weak or even absent, and the relative intensity of the \ion{[S}{2]} emission lines drop substantially \citep{pel12, wan19, ram20}. As discussed above, the galaxies in our sample show weak \ion{[S}{2]}$\lambda\lambda$6717,6731 nebular emission-lines relative to typical star-forming galaxies. They have high \ion{[N}{2]}/H$\alpha$ ratios consistent with $\gtrsim$ solar metallicity, but they exhibit anomalously weak \ion{[S}{2]} lines (see Fig.~\ref{fig:BPT_TotNar}), which could result from LyC photons escaping without encountering a low ionization outer edge of the nebula.

Similarly to O32, \ion{[S}{2]} deficiency does not appear to correlate strongly with the f$_{esc}$(LyC) of the known leakers. However, empirical correlations between line ratios and estimated f$_{esc}$(LyC) may be affected by geometric effects \citep{ste18, bas19, fle19} and shocks (see Section~\ref{section:composite}). Observations of individual local \ion{H}{2} regions show them to be geometrically complex, with significant spatial variation in oxygen line ratios, suggestive of regions from which LyC could escape \citep[e.g.,][]{zas11, wei15, keh16, kee17, mic18}. \ion{H}{2} regions may present channels carved into the ISM through which LyC flux could escape while other areas remain completely opaque to high-energy radiation.
Single-component (one-zone) photoionization models do not capture such complexity, failing to simultaneously reproduce the high and low ionization lines and escape fractions of LyC leakers. It has been shown that two-zone models, combining regions with a high- and low-ionization parameter, where one of which is density-bounded, do a better job at reproducing the observed line ratios, and f$_{esc}$(LyC) \citep[e.g.][]{ram20}. Predictions from the two-zone models in classical BPT diagrams vary with f$_{esc}$(LyC). LyC leakage does not influence \ion{[N}{2]} strongly as it originates from the highly excited region in the inner part of the \ion{H}{2} region, such that it remains unaffected when the edges of the \ion{H}{2} regions are trimmed. In contrast, the \ion{[S}{2]} lines are very sensitive to LyC leakage. The complexity of \ion{H}{2} regions could explain the variance in the O32 and \ion{[S}{2]} values displayed by the confirmed LyC leakers.

\begin{figure}[t!]
  \centering
   \includegraphics[width=0.95\columnwidth]{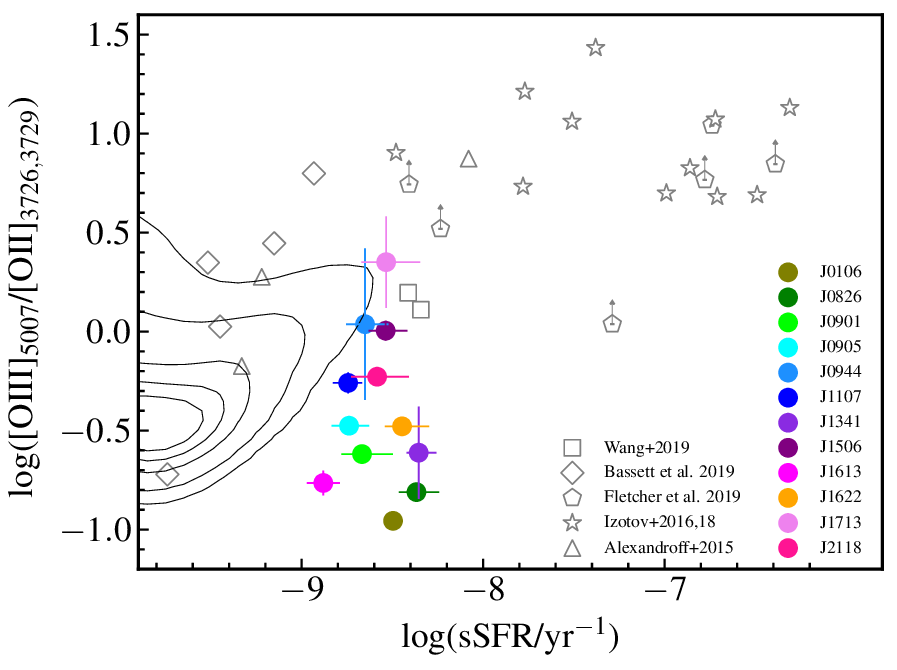}
  \caption{Total \ion{[O}{3]}$\lambda\lambda$5007 to \ion{[O}{2]}$\lambda\lambda$3726,3729 flux ratio compared to specific star formation rate. The grey contours show SDSS DR8 galaxies with contours at 25\%, 50\%, 75\%, 90\%, and 99\%. Black empty symbols are known Lyman continuum leaking galaxies: z$\sim$0.3 \ion{[S}{2]}-weak galaxies (squares; \citealp{wan19}), low redshift Green Pea galaxies (stars; \citealp{izo16b, izo16, izo18a, izo18b}), low redshift Lyman Break Analogs (triangles; \citealp{ale15}), z$\sim$3 star-forming galaxies (diamonds; \citealp{bas19}), and z$>$3 LACES galaxies (pentagons; \citealp{fle19}). Five targets from \citet{fle19} are not detected in \ion{[O}{2]}, the O32 values are 3$\sigma$ lower limits.}
  \label{fig:sSFR}
\end{figure}

It has been reported that the LyC emitters with the largest f$_{esc}$(LyC) also exhibit high sSFR ($\rm > 1\, Gyr^{-1}$; \citealp{bas19, kim20}). Additionally, hydrodynamical simulations find a correlation between increasing sSFR and increasing f$_{esc}$(LyC) \citep{yaj11, wis14}. In Fig.~\ref{fig:sSFR} we investigate the relationship between the O32 ratio and sSFR for our sample and the known LyC leakers shown in Fig.~\ref{fig:gal_prop}. 
The LyC leakers have a median sSFR of 10$^{-8.8}$ yr$^{-1}$, nearly 0.8 dex higher than the median value in our sample.  However, they span a wide range (3.4 dex) in sSFR, and the sSFRs values in our sample are similar to those of the most massive known LyC leakers.
The relatively lower sSFR values for our galaxies derive from their substantially higher M$_*$, $\sim$1.5 orders of magnitude higher than most known LyC leakers. 
Despite most of the confirmed LyC leakers show high sSFR, this may not be the relevant parameter for driving LyC leakage as the way in which they appear to be most distinct from other galaxies is not $\rm M_*$, but $\rm\Sigma_{SFR}$ (see Fig.~\ref{fig:gal_prop}). Similarly to our sample, almost all the individual observed LyC leakers to date show $\rm\Sigma_{SFR}$ higher than the average
$\rm\Sigma_{SFR}$ expected at their redshifts \citep{sharma16,nai20}.

An additional piece of evidence that a portion
of the LyC may be escaping the host galaxy comes from the emission and absorption line profiles. Contrary to typical starbursts, most of the galaxies in our sample lack \ion{Mg}{2} absorption near the systemic velocity, as seen in Section~\ref{subsection:kinematics}. This suggests that much of the photoelectric opacity to the LyC is in the tenuous wind itself rather than in the dense \ion{H}{2} regions. This is borne out by the nebular emission lines, which have usually broad line profiles, sometimes extending over the same range of velocities seen in absorption (see Fig.~\ref{fig:Velocities}).

Another intriguing line of evidence suggests that the galaxies in our sample may be leaking LyC photons. They have weak nebular emission lines, while detailed stellar population synthesis modeling of their UV-optical spectra shows that many of the galaxies have young ionizing stellar populations ($<10$ Myr) that should be producing copious nebular emission. 
To illustrate this, we compiled a set of color-matched galaxies from the eBOSS sample \citep{daw16} for each of the galaxies in our sample. We selected galaxies with g-r and r-i within $\pm$0.6 mag and redshift within $\pm$0.05, resulting in $10-200$ comparison galaxies per source. We found that our galaxies have much lower H$\beta$ EWs than the color-matched eBOSS galaxies, 
with a median H$\beta$ EW of 6.7\AA \ in our sample and 35\AA \  in the comparison sample. 
We note that dust alone can not account for the lack of strong emission in our sources.
The apparent Balmer emission line deficit is not an artifact of differential dust attenuation: the $3-5$ Myr old stars producing the ionizing photons share the same attenuation as the nebular emission lines excited by these stars. 

However, there may be other ways to explain these  observations. A possible scenario is that substantial numbers of LyC photons are absorbed by dust before ionizing hydrogen. Some amount of dust absorption seems likely in our sources as WISE 22 $\mu$m imaging shows that the galaxies are luminous in the restframe mid-IR. However, these galaxies are luminous in the GALEX far-UV bands, and their SEDs suggest relatively modest attenuation (A$\rm_V \sim 0.43$). Thus a complex “picket
fence” ISM geometry may be likely, with some high attenuation sightlines and some holes enabling LyC escape. The high incidence of strong outflows detected in our sample may be responsible for such holes in the ISM.

In summary, the galaxies in our sample show multiple indirect indications that they might be leaking LyC photons. They are characterized by high $\rm\Sigma_{SFR}$ and ionization parameters traced by the O32 ratio in line with those of the most massive known LyC leakers. Moreover, they lack of gas near zero velocity, and exhibit Balmer emission lines weaker than expected from stellar population synthesis modeling of their UV-optical spectra. Lastly, they show anomalously weak \ion{[S}{2]} lines. 
As our galaxies differ in many respects from known LyC leakers, our sample may offer an ideal opportunity to test what physical property is most closely linked to LyC escape. 
Directly measuring the LyC and determining f$_{esc}$(LyC) for our sample are necessary steps to confirm our hypothesis of potential LyC leakage. If our sample is found to have a significant f$_{esc}$(LyC), this would reveal that the LyC leakage process is not exclusively driven by low mass ($<$ 10$^8$ M$_{\odot}$) galaxies. Interestingly, the recent model by \citet{nai20} suggests that $<5$ \% of bright ($\rm M_{UV} < 18$) galaxies with log(M$_*$/M$_{\odot}$) $>$ 8 could account for $>80$\% of the reionization budget, making our sample 
potential analogs to the high redshift sources driving the reionization.

\section{Summary and Conclusions}
\label{section:Conclusions}

We use new optical and near-IR spectroscopy of 14 compact starburst galaxies at z $\sim$ 0.5, in combination with ancillary data, to study both the nature of their extreme ejective feedback episodes and the physical conditions in their dusty cores. These galaxies are massive ($\rm M_* \sim 10^{11} \,M_{\odot}$), compact (half-light radius $\sim$ few hundred pc), they have high star formation rates (mean $\rm SFR \sim 200 \, M_{\odot} yr^{-1}$) and star formation surface densities (mean $\rm \Sigma_{SFR} \sim 2000 \,M_{\odot} yr^{-1} kpc^{-2}$), and are known to exhibit extremely fast (mean maximum velocity $\sim -1890$ \kmps) outflows traced by \ion{Mg}{2} absorption lines (\citealp{tre07}; Davis et al., in prep.). 
 Our unique data set consists of a suite of both emission (\ion{[O}{2]}$\lambda\lambda$3726,3729, H$\beta$, \ion{[O}{3]}$\lambda\lambda$4959,5007, H$\alpha$, \ion{[N}{2]}$\lambda\lambda$6549, 6585, and \ion{[S}{2]}$\lambda\lambda$6716,6731) and absorption lines (\ion{Mg}{2}$\lambda\lambda$2796,2803, and \ion{Fe}{2}$\lambda$2586) that allow us to study the kinematics of the cool gas phase (T $\sim$ 10$^4$ K). The high M$_*$, SFR, and $\rm \Sigma_{SFR}$ values of these galaxies allow us to extend the dynamic range over which to investigate trends of outflow characteristics with galaxy properties. Our main conclusions are as follows:

1) The emission lines in 12/14 galaxies in our sample show a broad and blueshifted component. The \ion{[O}{2]} and H$\alpha$ broad emission lines exhibit average widths ($\sigma$) of 668 and 467 \kmps, and offsets of their central velocities from the systemic redshift (v$_{off}$) of 352 and 143 \kmps, respectively (Section~\ref{subsection:kinematics} and Fig.~\ref{fig:Fits}). Such line broadening and blueshifts clearly trace high velocity outflows. 

2) The ions studied in this work allow us to probe outflowing gas at different densities and distances from the central starburst. Absorption lines are sensitive to lower density gas and in our sample typically display somewhat higher maximum velocities than the emission lines  (Section~\ref{subsection:kinematics},  Fig.~\ref{fig:Velocities} and ~\ref{fig:comparison_vel}). This could reflect that the fastest outflowing gas has lower density, on average, which may be easier to accelerate. 

3) We characterize the physical conditions of the compact starburst using an ensemble of line ratio diagrams as key diagnostics of electron density, metallicity, and gas ionization. Our sample exhibits high electron density with a median value of 530 cm$^{-3}$ (Section~\ref{subsection:electron_density} and Fig.~\ref{fig:elecdens}), solar or super-solar metallicity, and a wide range of ionization parameter probed by the O32 ratio ranging from 0.11 to 2.24 (Section~\ref{subsection:ion_met} and Fig.~\ref{fig:O32_plot}). Our results show that the detected fast winds are most likely driven by stellar feedback resulting from the extreme central starburst, rather than by a buried AGN (Sections~\ref{section:AGN} and \ref{section:composite}).

4) We present multiple intriguing observational signatures that suggest that these galaxies may have substantial LyC photon leakage (Section~\ref{section:LyC_leak}). They have high $\rm\Sigma_{SFR}$ and ionization parameters comparable to those of the most massive known LyC leakers, as traced by the O32 ratio. They also lack gas in absorption near the systemic redshift and exhibit relatively weak Balmer emission lines. Finally, they show remarkably weak \ion{[S}{2]} lines compared to normal star-forming galaxies. As our galaxies are distinct from known LyC leakers in many regards (e.g., M$_*$ and sSFR), this sample presents an excellent chance to isolate which physical properties are most closely connected to LyC escape.

The compact starburst galaxies in our sample provide a unique opportunity to study star formation and feedback at its most extreme.  In a related paper we find that these galaxies are likely observed during a short-lived but potentially key phase of massive galaxy evolution (Whalen et al., in preparation). They have $\rm\Sigma_{SFR}$ values approaching the Eddington limit associated with stellar radiation pressure feedback \citep{tho05} and much of their gas may be violently blown out by powerful outflows that open up channels for LyC photons to escape. 

In a series of forthcoming papers based on high-resolution Keck/HIRES and integral field unit Keck/KCWI spectra, we will focus on deriving robust measurements of the physical properties, morphology, and extent of the galactic outflows in our sample. 
Such data on these unique galaxies provide strong observational constraints to theoretical simulations that aim to produce realistic galactic outflows. 
The comparison of outflow characteristics between simulations and observations will advance our understanding of  galactic feedback, particularly from stellar processes, during a crucial phase of massive galaxy evolution.

\section*{Acknowledgements}

We thank the referee for her/his time to provide a constructive report. The referee's thoughtful comments have helped to improve the clarity of the manuscript.
We acknowledge support from the National Science Foundation (NSF) under a collaborative grant (AST-1813299, 1813365, 1814233, 1813702, and 1814159) and from the Heising-Simons Foundation grant 2019-1659. S.~P. and A.~L.~C. acknowledge support from the Ingrid and Joseph W. Hibben endowed chair at UC San Diego. The data presented herein were obtained at the W. M. Keck Observatory, which is operated as a scientific partnership among the California Institute of Technology, the University of California and the National Aeronautics and Space Administration. The Observatory was made possible by the generous financial support of the W. M. Keck Foundation.
The authors wish to recognize and acknowledge the very significant cultural role and reverence that the summit of Maunakea has always had within the indigenous Hawaiian community.  We are most fortunate to have the opportunity to conduct observations from this mountain.

\bibliographystyle{aasjournal}
\bibliography{biblio}
\end{document}